\documentclass[prd,%preprint,
,superscriptaddress,%showpacs,
nofootinbib,%
tightenlines ]{revtex4}
\usepackage{epsfig}
\usepackage[colorlinks=true,linkcolor=blue,urlcolor=blue,citecolor=blue]{hyperref}
\usepackage{amsmath}
\usepackage{amsfonts}
\usepackage{amssymb}

\newcommand{\ben}{\begin{displaymath}}
\newcommand{\een}{\end{displaymath}}
\newcommand{\be}{\begin{equation}}
\newcommand{\ee}{\end{equation}}
\newcommand{\bea}{\begin{eqnarray}}
\newcommand{\eea}{\end{eqnarray}}

\newcommand{\nn}{\nonumber \\ }

\begin{document}
%\preprint{MKPH-T-06-16}
\title{Electromagnetic and gravitational local spatial densities for spin-1 systems}
 \author{J.~Yu.~Panteleeva}
  \affiliation{Institut f\"ur Theoretische Physik II, Ruhr-Universit\"at Bochum,  D-44780 Bochum,
 Germany}
\author{E.~Epelbaum}
 \affiliation{Institut f\"ur Theoretische Physik II, Ruhr-Universit\"at Bochum,  D-44780 Bochum,
 Germany}
\author{J.~Gegelia}
 \affiliation{Institut f\"ur Theoretische Physik II, Ruhr-Universit\"at Bochum,  D-44780 Bochum,
 Germany}
 \affiliation{Tbilisi State
University, 0186 Tbilisi, Georgia}
\author{U.-G.~Mei\ss ner}
 \affiliation{Helmholtz-Institut f\"ur Strahlen- und Kernphysik and Bethe
   Center for Theoretical Physics, Universit\"at Bonn, D-53115 Bonn, Germany}
 \affiliation{Institute for Advanced Simulation, Institut f\"ur Kernphysik
   and J\"ulich Center for Hadron Physics, Forschungszentrum J\"ulich, D-52425 J\"ulich,
Germany}
\affiliation{Tbilisi State  University,  0186 Tbilisi,
 Georgia}

\date{2 May 2023}
\begin{abstract}
 The matrix elements of the electromagnetic current and the energy-momentum tensor for sharply localized states of spin-1 systems
  % in zero averaged momentum as well as in moving frames
  are considered. 
Their interpretation as local spatial densities of various characteristics of the considered system is discussed. 
 
\end{abstract}

\maketitle

\section{Introduction}

Since the development of quantum mechanics it is well known that classical physics is not adequate for describing atomic and subatomic objects. Still, our intuition and the 
language are so strongly dominated by the classical picture of the world, that we often trade rigorous mathematical expressions for less accurate but better understandable 
concepts. The charge density of the nucleon serves as a good example. While hadrons certainly possess complicated electromagnetic properties, low-energy electron-hadron scattering can be well described utilizing the one-photon-exchange approximation parameterized in terms of electromagnetic form factors. 
Motivated by this approximation,  three-dimensional  Fourier transforms of  the form factors in the Breit frame are often interpreted as spatial densities of the corresponding hadrons. This picture fits well to our classical intuition. 
It originates from the seminal papers on electron-proton scattering by Hofstadter, Sachs and others in the 60ties of the last century \cite{Hofstadter:1958,Ernst:1960zza,Sachs:1962zzc}.  
Similar interpretations have also been proposed for the Fourier transforms of the gravitational form factors and for various local distributions  \cite{Polyakov:2002wz,Polyakov:2002yz,Polyakov:2018zvc}.  
While the classical analogy implies that, e.g., electromagnetic 
properties of the nucleon can,  to some extent,  be described by the charge and magnetization densities, in reality there is no ``true charge density" 
which characterizes the actual distribution of the charge ``inside" the nucleon. In this sense the spatial densities depend on the adopted definition. 
It has been repeatedly pointed out that the identification of spatial density distributions with the Fourier transforms of the corresponding form factors in the Breit frame suffers from conceptual problems \cite{Burkardt:2000za,Miller:2007uy,Miller:2009qu,Miller:2010nz,Jaffe:2020ebz,Miller:2018ybm,Freese:2021czn}.   
In Ref.~\cite{Jaffe:2020ebz}, it was shown on the example of a spin-$0$ system that the expression for the charge density in terms of the Breit frame distribution follows only in the static limit of an infinitely heavy particle. 

The issue of a proper definition of the spatial distributions of matrix elements of local operators has attracted much attention in the last few years.  For example, the light-front approach allows one to define purely intrinsic spatial densities, which have probabilistic interpretation \cite{Burkardt:2000za, Miller:2007uy, Miller:2009qu, Miller:2010nz, Guo:2021aik,Freese:2023jcp}, however, the corresponding densities are obtained only as two-dimensional distributions. 
The relationship between these densities and the non-relativistic three-dimensional distributions in the Breit frame in terms of the Abel transforms was studied in Refs.~\cite{Panteleeva:2021iip,Freese:2021mzg, Kim:2021jjf,Kim:2021kum,Kim:2022bia,Kim:2022syw}.
 Alternatively, the phase-space approach \cite{Lorce:2020onh,Lorce:2022cle,Lorce:2018egm,Chen:2022smg,Chen:2023dxp,Hong:2023tkv} allows one to define fully relativistic and unambiguous spatial densities, which in contrast to the light-front ones are
three-dimensional. However, these densities do not possess a strict probabilistic interpretation. 

A proper definition of the three-dimensional charge density by using sharply localized states has been revisited for a spin-$0$ system in Ref.~\cite{Epelbaum:2022fjc}. It turned out that the same definition was actually suggested long ago in the largely overlooked work by Fleming in Ref.~\cite{Fleming:1974af}. In Ref.~\cite{Epelbaum:2022fjc}, closely following 
the logic of Ref.~\cite{Jaffe:2020ebz}, the charge density possessing the usual probabilistic interpretation  has been defined in the zero average momentum 
frame (ZAMF) of the system as well as 
in moving frames by using spherically symmetric sharply localized wave packets.\footnote{The ZAMF is defined as a Lorentz frame with the vanishing expectation value of the three-momentum for the state, specified by a spherically symmetric packet. 
For wave packets with a sharp localization around an eigenstate of the four-momentum operator, the ZAMF coincides with the rest-frame of the system.} 
This definition has also been generalized to spin-1/2 and spin-3/2 systems and to the gravitational densities \cite{Panteleeva:2022khw,Panteleeva:2022uii,Alharazin:2022xvp,Carlson:2022eps}.  

The aim of the current paper is to work out the details of the novel definition for spin-1 systems for the electromagnetic as well as the gravitational local spatial densities.  
The electromagnetic densities of spin-1 systems have attracted much attention. In Ref.~\cite{Lorce:2022jyi}  the relativistic 2D charge densities of the deuteron and their frame dependence have been studied in the phase-space approach with the result that less frame dependence compared to the case of the spin-1/2 systems has been found. 
%{\color{red}Although the phase-space approach does not have a strict probabilistic interpretation, it provides access to the Breit-frame  as well %as to the light-front picture with the conceived (XXX what does that mean?XXX)  interpretation. I BELIEVE THIS SENTENCE IS BORROWED %FROM LORCE'S PAPER, WHY DO NOT WE JUST DROP IT?}
In Ref.~\cite{Kim:2022bia} the relation between the 3D and 2D  Breit-frame expressions and the 2D infinite-momentum frame (IMF) charge densities have been investigated using the definition of the Wigner distributions and the Abel transformation.  The densities were expressed in terms of  multipole expansions which provide a more clear physical meaning than the helicity-amplitudes of the form factors usually used \cite{Carlson:2009ovh,Alexandrou:2009hs,Carlson:2007xd,Lorce:2022jyi}.

The spatial gravitational densities for spin-1 systems have been also extensively discussed in recent years. In particular, in Ref.~\cite{Polyakov:2019lbq} the multipole expansion of the gravitational densities for spin-1 systems was suggested and computed in the Breit frame. At the same time, important properties of the EMT of  spin-1 systems were derived and another parametrization of this quantity was suggested in Ref.~\cite{Cosyn:2019aio}. Further, the multipole expansion of the densities for the $\rho$-meson in the light-cone quark model was studied in Ref.~\cite{Sun:2020wfo}.  
Recently the two-dimensional light front densities of spin-1 systems were calculated and discussed in Ref.~\cite{Freese:2022yur}.

In this paper we express the spatial densities of spin-1 systems in terms of the multipole expansion. 
Analogously to other cases, we consider sharply localized  and \textit{spherically symmetric} wave packets and obtain local spatial distributions for the ZAMF and moving frames, as well as traditional distributions for the Breit frame.

Our work is organized as follows. In Sect.~\ref{DefSLS} we specify the details of the localized states used in the definitions of local spatial densities. 
In Section~\ref{ElDen} we define the electromagnetic densities corresponding to the matrix elements of the electromagnetic current  in the ZAMF and discuss the static approximation. Gravitational spatial densities of the EMT  operator  in the ZAMF and the static approximation are 
considered in Section~\ref{GrD}. In Sect.~\ref{movingFs} we obtain the expressions of spatial densities in moving frames, and Sect.~\ref{summary} contains our summary.

\section{Sharply localized states}
\label{DefSLS}

We are interested in matrix elements of the electromagnetic current and the EMT operators in spatially localized 
normalizable Heisenberg-picture states. Such states can be specified in terms of wave packets 
\begin{equation}
|\Phi, {\bf X},\sigma \rangle = \int \frac{d^3 {p}}{\sqrt{2 E (2\pi)^3}}  \, \phi(\sigma,{\bf p}) \, e^{-i {\bf p}\cdot{\bf X}} |p ,\sigma \rangle\,,
\label{statedefN2}
\end{equation}
with the eigenstates of the four-momentum $|p,\sigma\rangle$, 
characterizing our spin-1 system with momentum $p$ and polarization $\sigma$, normalized as
\begin{equation}
\langle p',\sigma'|p,\sigma\rangle = 2 E (2\pi )^3 \delta_{\sigma'\sigma}\delta^{(3)} ({\bf p'}-{\bf p})\,.
\label{NormStateN}
\end{equation}
Here,  $p=(E,{\bf p})$, $E=\sqrt{m^2+{\bf p}^2}$ and $m$ is the mass of the system. 
The spatial translation vectors ${\bf X}$ can be interpreted as the position of the electromagnetic or the gravitational center of the system depending 
on the considered distributions, see  Refs.~\cite{Epelbaum:2022fjc,Panteleeva:2022khw,Panteleeva:2022uii}. It follows from the normalization of the wave packet that 
the profile function satisfies the condition
\begin{equation}
\int d^3 {p} \,  | \phi(\sigma,{\bf p})|^2 =1\,.  
\label{normN}
\end{equation}

To {\it define} the spatial density distributions of a physical system we use spherically symmetric wave
packets with profile functions  $\phi(\sigma, {\bf p}) = \phi({\bf p}) = \phi(|{\bf p} |)$ that are also spin-independent in case of systems of non-zero spin.
The average momentum of the system in the state specified by such a packet is equal to zero. Therefore, we identify the corresponding density distributions as 
characterizing the system in the ZAMF.
 
 For our calculations below it is convenient to define dimensionless profile functions 
\begin{equation}
\phi({\bf p}) = R^{3/2} \, \tilde \phi(R  {\bf p})\,,
\label{packageFormN}
\end{equation} 
where $R$ specifies the size of the wave packet with small values of $R$ corresponding to sharp localization.

\section{Electromagnetic densities}
\label{ElDen}

The matrix element of the electromagnetic  current operator for a spin-1 system for momentum eigenstates can be parameterized in terms of three form-factors \cite{Arnold1980}
\be
\langle p',\sigma'|\hat{j}^{\mu}({\bf r},0)|p,\sigma\rangle =-e^{-i({\bf p}'-{\bf p})\cdot {\bf r}}\epsilon^{\star}_{\alpha}(p',\sigma')\epsilon_\beta(p,\sigma)\left[2P^\mu g^{\alpha\beta}G_1(q^2)+(q^\alpha g^{\mu\beta}-q^\beta g^{\mu\alpha})G_2(q^2)-P^\mu q^\alpha q^\beta \frac{G_3(q^2)}{M^2}\right],
\label{Jmupar}
\ee
where $q=p'-p$ and $M$ is an arbitrary mass parameter, which is introduced to make the form factors dimensionless. 
It is natural to take $M$ equal to the physical mass $m$ of the system. However, to avoid the mixing of terms 
of different orders of $1/m$, when considering the static limit below, it is important to distinguish between 
$m$ and $M$. 
Therefore we put the parameter $M$ equal to $m$ only at the end of calculations, i.e. after performing the systematic expansion in $1/m$, whenever applicable. In  appendix~\ref{dist}, an explicit example is given to further corroborate
this issue.
The polarization 4-vectors in Eq.~(\ref{Jmupar}) are defined in standard way \cite{Varshalovich}:
\be
\epsilon^\mu(p,\sigma)=\left(\dfrac{{\bf p}\cdot{\boldsymbol{\hat{\epsilon}_\sigma}}}{m},{\boldsymbol{\hat{\epsilon}_\sigma}}+\dfrac{{\bf p}\cdot{\boldsymbol{\hat{\epsilon}_\sigma}}}{m(m+E)}{\bf p}\right),
\ee
where $\sigma\in\{+,-,0\}$ and the three-dimensional polarization basis vectors in the spherical representation 
are given by
\be
{\boldsymbol{\hat{\epsilon}_{\pm}}}=\mp\dfrac{1}{\sqrt{2}}(1,\pm i,0),\ \ {\boldsymbol{\hat{\epsilon}_{0}}}=(0,0,1).
\label{PolarVect}
\ee

 \medskip
 
The matrix element of the electromagnetic current operator for the state defined in Eq.~(\ref{statedefN2}) takes the following form
\begin{eqnarray}
  j^\mu_\phi({\bf r}) &\equiv & \langle \Phi, {\bf X},\sigma' | \hat j^\mu ({\bf x}, 0 ) | \Phi, {\bf X},\sigma \rangle  \nn
                                &=&- \int \frac{d^3 {P} \, d^3 {q}}{(2\pi)^3 \sqrt{4 E E'}} \, \epsilon^{\star}_{\alpha}(p',\sigma')\epsilon_\beta(p,\sigma)
\left[2P^\mu g^{\alpha\beta}G_1(q^2)+(q^\alpha g^{\mu\beta}-q^\beta g^{\mu\alpha})G_2(q^2)-P^\mu q^\alpha q^\beta \frac{G_3(q^2)}{M^2}\right] \nn
                &\times& \phi\bigg( {\bf P} -
\frac{\bf q}{2}\bigg) \, \phi^\star\bigg( {\bf P} +\frac{\bf q}{2}\bigg)  \, e^{ - i {\bf q}\cdot {\bf  r}} ,
\label{rhoint2N}
\end{eqnarray}  
where ${\bf P}=({\bf p}' + {\bf p})/2$, $ {\bf q}={\bf p}' - {\bf p}$, $E=\sqrt{m^2+ {\bf P}^2 - {\bf P}\cdot {\bf q} +{\bf q}^2/4 } $,
 $E'=\sqrt{m^2+ {\bf P}^2 + {\bf P}\cdot {\bf q} +{\bf q}^2/4 } $ and ${\bf  r} = {\bf  x}-{\bf  X}$.

\subsection{Electromagnetic densities in the ZAMF}

To obtain the electromagnetic spatial densities corresponding to internal structure of a spin-1 system we consider sharply localized wave packets in Eq.~(\ref{rhoint2N}).  
Using the method of dimensional counting of Ref.~\cite{Gegelia:1994zz} 
for the form factors $G_1(q^2)$, $G_2(q^2)$ and $G_3(q^2)$ decaying  for large $q^2$ as $1/q^4$, $1/q^4$ and $1/q^6$ (or faster), respectively, the $R\to 0$ limit in Eq.~(\ref{rhoint2N}) can be taken as discussed in
Ref.~\cite{Epelbaum:2022fjc}.
The final result for spherically symmetric wave packets with $\phi( {\bf {
  P}}) =  \phi(| {\bf {
  P}}|) $, takes the form
\bea
j^0({\bf r}) & = & \int \frac{d^2\hat{n}}{4\pi} \dfrac{d^3 {q}}{(2\pi)^3} \Bigg\{
\delta_{\sigma'\sigma}\mathcal{G}_0(-{\bf q}_\perp^2)+\hat{Q}^{kl}_{\sigma'\sigma}\hat{n}^k\hat{n}^l\dfrac{{\bf q}_\perp^2}{2m^2}\mathcal{G}_1(-{\bf q}_\perp^2)+\hat{Q}^{kl}_{\sigma'\sigma}q_\perp^k q_\perp^l\dfrac{\mathcal{G}_2(-{\bf q}_\perp^2)}{2m^2}\Bigg\} e^{ - i {\bf q}\cdot {\bf  r}}, \nn
{\bf j}({\bf r}) & =&\dfrac{1}{m}\int \frac{ d^2\hat{n} }{4\pi} \frac{ d^3 {q}}{(2\pi)^3} \,  {\bf \hat{n}} \
{\bf \hat{n}}\cdot(i{\bf \hat S}_{\sigma'\sigma}\times{\bf q})\mathcal{M}(-{\bf q}_\perp^2)  e^{ - i {\bf q}\cdot {\bf  r}},
\label{rhoint3Na}
\eea
where ${\bf\hat{S}}_{\sigma'\sigma}$ and $\hat{Q}^{kl}_{\sigma'\sigma}$ are the spin and the quadrupole operators, respectively, defined in Appendix~\ref{append}, and ${\bf\hat{ n}}$ is a three-dimensional unit vector. 
Here and in what follows, ${\bf a}_\parallel \equiv {\bf a} \cdot
{\bf \hat n} \, {\bf \hat n} $ and ${\bf a}_\perp \equiv {\bf a} -  {\bf a} \cdot {\bf \hat n} \, {\bf \hat n} 
$ denote the components of a
vector ${\bf a}$ parallel and perpendicular to the unit vector $\bf \hat n$, respectively, and
$a_\parallel \equiv | {\bf a}_\parallel |$,  $a_\perp \equiv | {\bf
  a}_\perp |$. 
  The spatial densities defined via Eq.~(\ref{rhoint3Na}) do not depend on the form of the radial profile function of the wave packet.  
The combinations of the form factors appearing in Eq.~(\ref{rhoint3Na}) are given by
\bea
\mathcal{G}_0(-{\bf q}_\perp^2)&=&G_1(-{\bf q}_\perp^2)\left(1-\dfrac{{\bf q}_\perp^2}{6m^2}\right)+G_2(-{\bf q}_\perp^2)\dfrac{{\bf q}_\perp^2}{6m^2}+G_3(-{\bf q}_\perp^2)\dfrac{{\bf q}_\perp^2}{6m^2}\left(1-\dfrac{{\bf q}_\perp^2}{4m^2}\right)\,,\nonumber\\
\mathcal{G}_1(-{\bf q}_\perp^2)&=&G_1(-{\bf q}_\perp^2)-G_2(-{\bf q}_\perp^2)+\dfrac{{\bf q}_\perp^2}{4m^2}G_3(-{\bf q}_\perp^2)\,,\nonumber\\
\mathcal{G}_2(-{\bf q}_\perp^2)&=&-G_3(-{\bf q}_\perp^2)\,,\nonumber\\
\mathcal{M}(-{\bf q}_\perp^2)&=&-G_1(-{\bf q}_\perp^2)+\dfrac{G_2(-{\bf q}_\perp^2)}{2}-\dfrac{{\bf q}_\perp^2}{4m^2}G_3(-{\bf q}_\perp^2)\,.
\label{ourGs}
\eea
The densities of Eq.~(\ref{rhoint3Na}) can be also parameterized in the following form 
\bea
j^0({\bf r})&=&\delta_{\sigma'\sigma}\rho_0(r)+\hat{Q}^{ij}_{\sigma'\sigma}Y_2^{ij}({\bf\hat{ r}})\rho_2(r),\nonumber\\
{\bf j}({\bf r})&=& {\bf \hat S}_{\sigma'\sigma}\times {\boldsymbol {Y}}_1({\bf\hat{ r}}) {{\rho}}_M(r) \,,
\eea   
where $Y_i({r})$ are multipoles defined in the Appendix~\ref{append} and %three dimensional electric monopole and quadrupole densities have the following form, which is different in compare to the densities in static approximation 
\bea
\rho_0(r)&=&\int \frac{d^2\hat n}{4\pi} \, \delta(r_\parallel) \tilde{\mathcal{G}}_0(r_\perp)  \,,\nonumber\\
\rho_2(r)&=& - \dfrac{1}{4}\int \frac{d^2\hat n}{4\pi} \, \delta(r_\parallel)\left[\left(3\dfrac{r_\parallel^2}{r^2}-1\right)  \dfrac{1}{m^2}\hat{O}_2(r_\perp)\tilde{\mathcal{G}}_1(r_\perp)
+\left(3\dfrac{r_\perp^2}{r^2}-1\right) \dfrac{1}{m^2}r^2_\perp\hat{O}_1(r_\perp)\tilde{\mathcal{G}}_2(r_\perp)
\right]\,, \nonumber\\
{\rho}_M(r)&=& - \frac{1}{2 m} \int \frac{d^2\hat{n}}{4\pi} \, \delta(r_\parallel) \, \frac{ {r_\perp}}{r}  \frac{d}{d r_\perp} \tilde{\mathcal{M}}(r_\perp) \,. %\nonumber\\
\eea
Here, the differential operators $\hat{O}_1(r_\perp)$ and $\hat{O}_2(r_\perp)$ are given by 
\bea\label{Ooperators}
\hat{O}_1(r_\perp)& =&\dfrac{1}{r_\perp}\dfrac{d}{dr_\perp}\dfrac{1}{r_\perp}\dfrac{d}{dr_\perp}\,,\nonumber\\
\hat{O}_2(r_\perp)& =&\dfrac{1}{r_\perp^2}\dfrac{d}{dr_\perp}r_\perp^2\dfrac{d}{dr_\perp}\,,
\eea
and we have introduced the two-dimensional Fourier transforms of the form factors
\bea
\tilde{\mathcal{G}}_i(r_\perp) &= &\int \dfrac{d^2q_{\perp}}{(2\pi)^2}e^{-i{\bf q}_{\perp}\cdot{\bf r}_{\perp}}\mathcal{G}_i(-{\bf q}_\perp^2)\,,\nonumber\\
\tilde{\mathcal{M}}(r_\perp) & = & \int \dfrac{d^2q_{\perp}}{(2\pi)^2}e^{-i{\bf q}_{\perp}\cdot{\bf r}_{\perp}}\mathcal{M}(-{\bf q}_\perp^2)\,.
\eea

\subsection{Electromagnetic densities in the Breit frame}
The traditional (``naive'') densities in terms of the Fourier transforms of the form factors in the Breit frame emerge by first expanding 
the integrand in Eq.~(\ref{rhoint2N}) in inverse powers of $m$ up to leading order prior to  performing the integration \cite{Miller:2018ybm,Jaffe:2020ebz} (notice that for this expansion it is 
important to distinguish between $m$ and $M$), and then expanding the integrands in powers of $R$ around $R = 0$ and keeping terms up to the zeroth order.  The resulting expressions read: 
\bea
j^0_{\text{naive}}({\bf r}) 
&=& \int \dfrac{d^3 q}{(2\pi)^3}e^{-i{\bf q}\cdot {\bf r}}\left(G_C(-{\bf q}^2)\delta_{\sigma\sigma'}+\dfrac{G_Q(-{\bf q}^2)}{2 m^2}\hat{Q}^{km}_{\sigma'\sigma}q^mq^k\right) 
 \equiv 
\delta_{\sigma\sigma'}\rho_C^{\text{naive}}(r)+\hat{Q}_{\sigma'\sigma}^{km}Y^{km}_2({ {\bf\hat{ r}}})\rho_Q^{\text{naive}}(r)\,,\nonumber\\
{\bf j}_{\text{naive}}({\bf r})&=& 
\int \dfrac{d^3 q}{(2\pi)^3}e^{-i{\bf q}\cdot {\bf r}}\dfrac{G_M(-{\bf q}^2)}{2m}i({\bf\hat{S}}_{\sigma'\sigma}\times {\bf q}) 
\equiv \dfrac{ ({\bf\hat{S}}_{\sigma'\sigma}\times {\boldsymbol \nabla})}{2m}\rho_M^{\text{naive}}(r)\,,
\label{staticEM}
\eea
where the electric monopole $G_C$, the electric quadrupole $G_Q$, and  the magnetic dipole $G_M$  form factors in the Breit frame are given by
\bea
G_C(-{\bf q}^2)&=&G_1(-{\bf q}^2)+\dfrac{{\bf q}^2}{6 m^2}G_3(-{\bf q}^2)\,,\nonumber\\
G_Q(-{\bf q}^2)&=&-G_3(-{\bf q}^2)\,, \nonumber\\
G_M(-{\bf q}^2)&=&G_2(-{\bf q}^2)\,.
\label{BFGs}
\eea
These expressions coincide with the traditional expressions for the current densities of a spin-1 system obtained in the Breit frame, see for example Ref.~\cite{Lorce:2022jyi}, after  expanding the 
latter
in inverse powers of $m$ and keeping the leading-order terms.\footnote{Notice that when expanding the densities from Ref.~\cite{Lorce:2022jyi} in inverse powers of $m$, one has  to divide the expressions from that work by a factor of $2 m$, take into account that the mass parameter $M$ in the parameterization of the form factors is not the mass in which the expansion is done, and keep only the leading-order term for each component of the current separately.}
The electric charge density distribution $\rho_C^{\text{naive}}(r)$, the electric quadrupole  charge distribution $\rho_Q^{\text{naive}}(r)$, 
and the magnetic density  $\rho_M^{\text{naive}}(r)$  have the following form 
 \bea
 \rho_C^{\text{naive}}(r)&=&\int \dfrac{d^3 q}{(2\pi)^3}e^{-i{\bf q}\cdot{\bf r}}G_C(-{\bf q}^2)\,,\\
 \rho_Q^{\text{naive}}(r)&=&-\dfrac{1}{2 m^2}r\dfrac{d}{dr}\dfrac{1}{r}\dfrac{d}{dr}\int \dfrac{d^3 q}{(2\pi)^3}e^{-i{\bf q}\cdot{\bf r}}G_Q(-{\bf q}^2)\,,\\
  \rho_M^{\text{naive}}(r)&=&\int \dfrac{d^3 q}{(2\pi)^3}e^{-i{\bf q}\cdot {\bf r}}G_M(-{\bf q}^2)\,.
 \eea
 As it was already discussed in Refs.~\cite{Jaffe:2020ebz,Epelbaum:2022fjc}, these densities describe the leading-order approximation to the matrix element of the current operator of systems
 %with intrinsic sizes
 % encoded in form factors are
 % much larger than the Compton wavelength $1/m$
in a state with localization much larger than the Compton wavelength $1/m$ yet much smaller than all intrinsic scales encoded in form factors. Clearly, for light hadrons with the intrinsic size being smaller than or comparable to the Compton wavelength, such an approximation becomes invalid.

\section{Gravitational densities}
\label{GrD}

Next we consider the local spatial densities corresponding to the matrix elements of the EMT operator. 
As emphasized in Ref.~\cite{Panteleeva:2022uii}  these densities differ significantly from the ones of the electromagnetic current. 
This is due to the fact that  a superposition of eigenstates of the electric 
charge operator, which makes the localized packet, is again an eigenstate of the charge operator with the same eigenvalue, while this is not the case for the energy-momentum operator.  

The matrix elements of the EMT of a spin-1 system in one-particle eigenstates of the energy-momentum operator can be parametrized in terms of form factors as follows  \cite{Polyakov:2019lbq}
\bea
&&\langle p', \sigma'| \hat{T}_{\mu\nu}({\bf x} ,0)| p,\sigma \rangle =\epsilon^{\star\beta}(p',\sigma')\epsilon^\alpha(p,\sigma) e^{-i  {\bf q} \cdot {\bf x}}  \Bigg[ 2P_\mu P_\nu \left(-g_{\alpha\beta}A_{0}(q^2)+\dfrac{P_\alpha P_\beta}{M^2}A_1(q^2)\right) \nonumber\\
&&\qquad + 2\big(P_\mu\left[g_{\nu\beta}P_\alpha+g_{\nu\alpha} P_\beta\right]+P_\nu\left[g_{\mu\beta}P_\alpha+g_{\mu\alpha} P_\beta\right]\big)J(q^2) + \dfrac{1}{2}\left(q_\mu q_\nu-g_{\mu\nu} q^2\right)\left(g_{\alpha\beta}D_0(q^2)+\dfrac{P_\alpha P_\beta}{M^2}D_1(q^2)\right) \nonumber\\
&&\qquad +\Big[\dfrac{1}{2}q^2\left(g_{\mu\alpha}g_{\nu\beta}+g_{\mu\beta}g_{\nu\alpha}\right)-\left(g_{\nu\beta}q_\mu+g_{\mu\beta}q_\nu\right)P_\alpha+\left(g_{\nu\alpha}q_\mu+g_{\mu\alpha}q_\nu\right)P_\beta-4g_{\mu\nu}P_\alpha P_\beta\Big]E(q^2) \nonumber\\
&&\qquad +\left(g_{\mu\alpha}g_{\nu\beta}+g_{\mu\beta}g_{\nu\alpha}-\dfrac{1}{2}g_{\mu\nu}g_{\alpha\beta}\right)M^2\overline{f}(q^2)+g_{\mu\nu}\left(g_{\alpha\beta}M^2\overline{c}_0(q^2)+P_\alpha P_\beta\, \overline{c}_1(q^2)\right)\Bigg]\,.
\label{EMTdef}
\eea
Here, we again distinguish between the mass of the system $m$ and the mass parameter $M$, which can be absorbed in the normalization of the form factors. 
Notice that in the parametrization we also included the non-conserved part of the EMT (namely the form factors $\bar f(q^2),\ \bar c_0(q^2)$ and $\bar c_1(q^2)$), so that e.g.~in QCD, one can consider the quark and gluon EMTs separately. However, for a conserved EMT these form factors vanish.

To define  the spatial densities associated with the EMT we consider its matrix element in a state specified by Eq.~(\ref{statedefN2})  and take the limit of sharply localized states. 
The considered matrix element of the EMT operator has the form
\bea
t^{\mu\nu}_{\phi}( {\bf r}) &=& \langle \Phi, {\bf X} |T^{\mu\nu}({\bf x},0)| \Phi, {\bf X} \rangle = \int\dfrac{d^3 {P} d^3 {q}}{(2\pi)^3 \sqrt{4 E' E }}  \, \phi^\star({\bf p'}) \,  \phi({\bf p})  e^{i {\bf q}\cdot{\bf X}}
\langle p' |T^{\mu\nu}({\bf x},0) |p \rangle \nonumber\\
& =& \int \dfrac{d^3 {P} \, d^3 q}{(2\pi)^3 \sqrt{4 E
    E'}}\, \phi\bigg({\bf P} -
\frac{\bf q}{2}\bigg) \, \phi^\star\bigg({\bf P} +\frac{\bf q}{2}\bigg)  \, e^{-i {\bf q}\cdot {\bf  r}} \epsilon^{\star\beta}(p',\sigma')\epsilon^\alpha(p,\sigma) \Bigg[ 2P_\mu P_\nu \left(-g_{\alpha\beta}A_0(q^2)+\dfrac{P_\alpha P_\beta}{M^2}A_1(q^2)\right) \nonumber\\
&+ &2\big(P_\mu\left[g_{\nu\beta}P_\alpha+g_{\nu\alpha} P_\beta\right]+P_\nu\left[g_{\mu\beta}P_\alpha+g_{\mu\alpha} P_\beta\right]\big)J(q^2) + \dfrac{1}{2}\left(q_\mu q_\nu-g_{\mu\nu} q^2\right)\left(g_{\alpha\beta}D_0(q^2)+\dfrac{P_\alpha P_\beta}{M^2}D_1(q^2)\right) \nonumber\\
&+&\Big[\dfrac{1}{2}q^2\left(g_{\mu\alpha}g_{\nu\beta}+g_{\mu\beta}g_{\nu\alpha}\right)-\left(g_{\nu\beta}q_\mu+g_{\mu\beta}q_\nu\right)P_\alpha+\left(g_{\nu\alpha}q_\mu+g_{\mu\alpha}q_\nu\right)P_\beta-4g_{\mu\nu}P_\alpha P_\beta\Big]E(q^2) \nonumber\\
&+&\left(g_{\mu\alpha}g_{\nu\beta}+g_{\mu\beta}g_{\nu\alpha}-\dfrac{1}{2}g_{\mu\nu}g_{\alpha\beta}\right)M^2\overline{f}(q^2)+g_{\mu\nu}\left(g_{\alpha\beta}M^2\overline{c}_0(q^2)+P_\alpha P_\beta\overline{c}_1(q^2)\right)\Bigg] % \nn & 
       % \nn &\times
               \,.
\label{rhoint2G}
\eea
%where for momentum variables we use notations of the p.

\subsection{Gravitational densities in the ZAMF}

Analogously to the case of the electromagnetic current we take the limit of sharply localized packets by applying the method of dimensional counting of Ref.~\cite{Gegelia:1994zz}. 
However, when expanding in powers of $R$ around $R=0$, we now keep explicitly only the leading-order terms for each form factor separately and denote by ``Rest" all other contributions. This is because different parts of the EMT require a different
physical interpretation
as discussed in Refs.~\cite{Freese:2021mzg,Freese:2022fat,Panteleeva:2022uii}. For the form factors decaying for large $q^2$ as $A_0(q^2)\sim 1/q^4$, $A_1(q^2)\sim 1/q^6$, $J(q^2)\sim 1/q^4$, $D_0(q^2)\sim 1/q^6$, $D_1(q^2)\sim 1/q^8$, $E(q^2)\sim 1/q^4$,  $\bar{f}(q^2)\sim1/q^2$, 
$\bar{c}_0(q^2)\sim 1/q^4$ and $\bar{c}_1(q^2)\sim 1/q^6$ or faster, the final result reads 
\bea
t_{\phi}^{00}&=& N_{\phi, R}\int d^2\hat{n} \,\dfrac{d^3q}{(2\pi)^3}e^{-i{\bf q}\cdot{\bf r}}\Bigg\{\delta_{\sigma'\sigma}\mathcal{E}_0(-{\bf q}_\perp^2)+\hat{Q}^{kl}_{\sigma'\sigma}\hat{n}^k\hat{n}^l \dfrac{{\bf q}_\perp^2}{m^2}\mathcal{E}_1(-{\bf q}_\perp^2)+\dfrac{\mathcal{E}_2(-{\bf q}_\perp^2)}{m^2}\hat{Q}^{kl}_{\sigma'\sigma}q_\perp^k q^l_\perp
\Bigg\}\ + {\rm Rest} \,, \nonumber\\
t_{\phi}^{0i}&=&N_{\phi, R}\int d^2\hat{n} \,\dfrac{d^3q}{(2\pi)^3}\dfrac{ {\hat{n}^i }}{m}\mathcal{J}(-{\bf q}_\perp^2)
e^{-i{\bf q}\cdot{\bf r}} (i \, {\bf\hat{S}}_{\sigma'\sigma}\times   {\bf q})\cdot{\bf \hat n}\  + {\rm Rest}  \,,\nonumber\\
t_{\phi}^{ij}&=& N_{\phi, R}\int d^2\hat{n} \,\dfrac{d^3q}{(2\pi)^3}\hat{n}^i\hat{n}^j\Bigg\{\delta_{\sigma'\sigma}\mathcal{E}_0(-{\bf q}_\perp^2)
+\hat{Q}^{kl}_{\sigma'\sigma}\hat{n}^k\hat{n}^l\dfrac{{\bf q}^2_\perp}{m^2}\mathcal{E}_1(-{\bf q}_\perp^2)
+\dfrac{\mathcal{E}_2(-{\bf q}_\perp^2)}{m^2}\hat{Q}^{kl}_{\sigma'\sigma}q_\perp^kq_\perp^l\Bigg\} e^{-i{\bf q}\cdot{\bf r}}  \nonumber\\
&+&
N_{\phi, R,2}\int d^2\hat{n} \,\dfrac{d^3q}{(2\pi)^3}\Bigg\{\left(\delta_{ij}{\bf q}^2_{\perp}-q_iq_j\right)\Bigg[\delta_{\sigma'\sigma}\mathcal{D}_0(-{\bf q}_\perp^2)
+ \hat{Q}^{kl}_{\sigma'\sigma}\hat{n}^k\hat{n}^l\dfrac{{\bf q}^2_\perp}{m^2}\mathcal{D}_1(-{\bf q}_\perp^2)+\dfrac{\mathcal{D}_2(-{\bf q}_\perp^2)}{m^2}\hat{Q}^{kl}_{\sigma'\sigma} q_\perp^kq_\perp^l\Bigg]  \nonumber\\
&+& \delta_{ij}\Bigg[\delta_{\sigma'\sigma}m^2\mathcal{C}_0(-{\bf q}_\perp^2)+{\bf q}_\perp^2\hat{Q}^{kl}_{\sigma'\sigma}\hat{n}^k\hat{n}^l\mathcal{C}_1(-{\bf q}_\perp^2) 
+\mathcal{C}_2(-{\bf q}_\perp^2)\hat{Q}^{kl}_{\sigma'\sigma}q_\perp^kq_\perp^l\Bigg] 
\Bigg\}
e^{-i{\bf q}\cdot{\bf r}}\ + {\rm Rest} \,, 
\label{GSDZ}
\eea
where the explicit form of the linear combinations of the form factors, $\mathcal{E}_i(-{\bf q}_\perp^2)$,  $\mathcal{J}(-{\bf q}_\perp^2)$, $\mathcal{D}_i(-{\bf q}_\perp^2)$ and $\mathcal{C}_i(-{\bf q}_\perp^2)$ is specified in Appendix~\ref{linCofFFs}.
As mentioned above, we kept explicitly the leading-order contributions 
of the terms with the $\mathcal{D}_i(-{\bf q}_\perp^2)$ and  $\mathcal{C}_i(-{\bf q}_\perp^2)$ form factors, while the contributions of the same order (and lower) in $R$ stemming from the terms with the $\mathcal{E}_i(-{\bf q}_\perp^2)$ and $\mathcal{J}(-{\bf q}_\perp^2)$ form factors are not shown for the reason explained above.
% because of the separate physical interpretation of these contributions.
The spatial densities of Eq.~(\ref{GSDZ}) depend on the wave packet only via the overall normalization constants 
\bea
N_{\phi,R} &=& \frac{1}{R}  \int  \, d\tilde P \tilde P^3 |\tilde\phi({|\tilde
  {\bf P}|})|^2 
  \,,\nonumber\\
  N_{\phi,R,2} &=& \frac{R}{2}  \int  \, d\tilde P \tilde P |\tilde\phi({|\tilde
  {\bf P}|})|^2
  \,.
  \label{normaliz}
\eea
Notice that for $R\to 0$, the first normalization constant in Eq.~(\ref{normaliz}) goes to infinity while the second constant vanishes. 

\medskip

The energy distribution $t_{\phi}^{00}(r)$ can be written in the form of a three-dimensional multipole expansion as follows:
 \bea
t_{\phi}^{00}(r)=\rho_{E0}(r)\delta_{\sigma'\sigma}+\hat{Q}^{kl}_{\sigma'\sigma} Y_2^{kl}({\bf\hat{ r}}) \rho_{E2}(r)\,,
\eea
where the monopole and quadrupole energy distributions have the form 
\bea
\rho_{E0}(r)&=&N_{\phi,R}\int d^2\hat{n} \,\delta(r_\parallel)\varepsilon_0(r_\perp)\,,\nonumber\\
\rho_{E2}(r)&=&\dfrac{N_{\phi,R}}{2}\int d^2\hat{n} \,\delta(r_\parallel)\left[\left(3\dfrac{r_\parallel^2}{r^2}-1\right)\varepsilon_1(r_\perp)
+\left(3\dfrac{r_\perp^2}{r^2}-1\right)\varepsilon_2(r_\perp)\right]\,,
\eea
with
\bea
\varepsilon_0(r_\perp)&=&\tilde{\mathcal{E}}_0(r_\perp)\,,\nonumber\\
\varepsilon_1(r_\perp)&=&-\dfrac{1}{m^2}\hat{O}_2(r_\perp)\tilde{\mathcal{E}}_1(r_\perp)\,,\nonumber\\
\varepsilon_2(r_\perp)&=&-\dfrac{1}{m^2}r_\perp^2\hat{O}_1(r_\perp)\tilde{\mathcal{E}}_2(r_\perp)\,,\nonumber\\
\tilde{\mathcal{E}}_i(r_\perp)&=&\int \dfrac{d^2q_{\perp}}{(2\pi)^2}e^{-i{\bf q}_{\perp}\cdot{\bf r}_{\perp}}\mathcal{E}_i(-{\bf q}_\perp^2)\,,
\eea
where the differential operators $\hat{O}_i$ are defined in Eq.~\eqref{Ooperators}.

\medskip

The multipole expansion of the momentum-density distribution has the form
\be
t_\phi^{0i}(r)=\left({\bf\hat{S}}_{\sigma'\sigma}\times   {\bf Y}_1({\bf\hat{ r}})\right) \tilde J(r)\,,
\ee
where 
\bea
\tilde J(r) & = & \dfrac{N_{\phi,R}}{2}\int d^2\hat{n} \, \delta(r_\parallel) \frac{r_\perp}{r} \, J(r_\perp) \,,\nonumber\\
 J(r) & = & -\frac{1}{m} \frac{d}{d r_\perp} \, \tilde{\mathcal{J}}(r_\perp)\,,
\eea
with
\be
 \tilde{\mathcal{J}}(r_\perp)  =  \int \frac{d^2 q_\perp}{(2\pi)^2} \, e^{-i {\bf q}_\perp\cdot {\bf r}_\perp} {\mathcal{J}}(-{\bf q}_\perp^2)\,.
\ee

\medskip

The $ij$th components of the EMT can be written as the sum of three parts 
\be
t_{\phi}^{ij}(r)=t_0^{ij}(r)+t_2^{ij}(r)+t_3^{ij}(r)\,,
\ee
where the first term is called the flow tensor and has the form
\be
t_{0}^{ij}({\bf r})=N_{\phi, R}\int d^2\hat{n} \,\dfrac{d^3q}{(2\pi)^3}\hat{n}^i\hat{n}^j\Bigg\{\delta_{\sigma'\sigma}\mathcal{E}_0(-{\bf q}_\perp^2)+\hat{Q}^{kl}_{\sigma'\sigma}
\hat{n}^k\hat{n}^l\dfrac{{\bf q}^2_\perp}{m^2}\mathcal{E}_1(-{\bf q}_\perp^2) 
+\dfrac{\mathcal{E}_2(-{\bf q}_\perp^2)}{m^2}\hat{Q}^{kl}_{\sigma'\sigma} q_\perp^kq_\perp^l\Bigg\} e^{-i{\bf q}\cdot{\bf r}}\,.
\label{tij0def}
\ee 
After integrating over the momentum ${\bf q}$ and the unit vector  ${\bf \hat n}$ in Eq.~(\ref{tij0def}) we obtain the following expression
\bea
t_0^{ij}({\bf r})=
\delta_{\sigma'\sigma} \left( \delta^{ij} A_0(r) + Y_2^{ij}({\bf\hat{ r}}) \, B_0(r) \right) &+& \hat{Q}^{ij}_{\sigma'\sigma}  A_2(r)  
+2 \left(  \hat{Q}^{ik}_{\sigma'\sigma}  Y_2^{jk}({\bf\hat{ r}}) + \hat{Q}^{kj}_{\sigma'\sigma} Y_2^{ik}({\bf\hat{ r}})  -\delta_{ij} \hat{Q}^{kl}_{\sigma'\sigma} Y_2^{kl}({\bf\hat{ r}})  \right) B_2(r)  
  \nonumber\\
&+&Y_2^{kl}({\bf\hat{ r}})\hat{Q}^{kl}_{\sigma'\sigma}\left[\delta^{ij} \left( A_1(r) +\frac{1}{3} B_1(r) +2B_2(r)\right) + Y_2^{ij}({\bf\hat{ r}})  \, B_1(r)  \right] \,,
\eea
where
\bea
A_{0}(r)&=&\frac{N_{\phi,R}}{3} \int d^2\hat{n}  \,\delta(r_\parallel) 
\varepsilon_0(r_\perp),\nonumber\\
B_{0}(r)&=&\dfrac{N_{\phi,R}}{2}\int d^2\hat{n} \,\delta(r_\parallel) \left(3\dfrac{r_\parallel^2}{r^2}-1\right)\varepsilon_0(r_\perp) ,\nonumber\\
A_{1}(r)&=& N_{\phi,R}  \int d^2\hat{n} \,\delta(r_\parallel)\,\frac{r_\perp^4}{8r^4}\left[\left(\dfrac{4 r_\parallel^2}{r_\perp^2}-1\right)\varepsilon_1(r_\perp)
+  \left( 4 - \dfrac{r_\parallel^2}{r_\perp^2}\right)\varepsilon_2(r_\perp)\right],\nonumber\\
A_{2}(r)&=&2N_{\phi,R}  \int d^2\hat{n} \,\delta(r_\parallel) \,\frac{r_\perp^4}{8r^4}
\left[ \frac{1}{3}\left(\dfrac{8 r_\parallel^2}{r_\perp^2} + 1\right)  \varepsilon_1(r_\perp) - \dfrac{7 r_\parallel^2}{3 r_\perp^2} \, \varepsilon_2(r_\perp)\right],\nonumber\\
B_{1}(r)&=& N_{\phi,R}  \int d^2\hat{n} \,\delta(r_\parallel)\,\frac{r_\perp^4}{8r^4}\left[\left(\frac{35 r_\parallel^4}{r^4} + 3 - \frac{30 r^2_\perp}{r^2} \right)\varepsilon_1(r_\perp)
+\left( \frac{35 r_\parallel^2 r_\perp^2}{r^4} - 4 \right)\varepsilon_2(r_\perp)\right],\nonumber\\ 
B_{2}(r)&=& N_{\phi,R}  \int d^2\hat{n} \,\delta(r_\parallel)\,\frac{r_\perp^4}{8r^4}\left[\left(\dfrac{4 r_\parallel^2}{r_\perp^2}-1\right)\varepsilon_1(r_\perp)
- \dfrac{5 r_\parallel^2}{r_\perp^2} \, \varepsilon_2(r_\perp)\right].
\label{defsRhoJ}
\eea

The second part of $t_{\phi}^{ij}$ is the stress tensor, which describes the internal structure of the system  and has the form:
\be
t^{ij}_2({\bf r})=N_{\phi, R,2}\int d^2\hat{n} \,\dfrac{d^3q}{(2\pi)^3}\Bigg\{\left(\delta_{ij}{\bf q}^2_{\perp}-q_iq_j\right)\Bigg[\delta_{\sigma'\sigma}\mathcal{D}_0(-{\bf q}_\perp^2)+ \hat{Q}^{kl}_{\sigma'\sigma}\hat{n}^k\hat{n}^l\dfrac{{\bf q}^2_\perp}{m^2}\mathcal{D}_1(-{\bf q}_\perp^2)+\dfrac{\mathcal{D}_2(-{\bf q}_\perp^2)}{m^2}\hat{Q}^{kl}_{\sigma'\sigma} q_\perp^kq_\perp^l\Bigg]\Bigg\}e^{-i{\bf q}\cdot{\bf r}}.
\ee
It can be reduced to 
\bea
t_2^{ij}({\bf r})&=& N_{\phi, R,2}\int d^2\hat{n} \Bigg\{\delta_{\sigma'\sigma} \left[\delta^{ij}\hat{d}_1(r)+Y^{ij}_2({\bf\hat{ r}})\hat{d}_2(r)\right]
+  \hat{Q}^{kl}_{\sigma'\sigma} Y_2^{kl}({\bf\hat{ r}}) \left(\delta^{ij}  \hat{d}_3(r) + Y_2^{ij}({\bf\hat{ r}}) \hat{d}_4(r)\right) \nonumber\\
&+&
\hat{Q}^{ij}_{\sigma'\sigma} \hat{d}_5(r) 
+\left(\hat{Q}^{ik}_{\sigma'\sigma} Y_2^{j k}({\bf\hat{ r}}) + \hat{Q}^{jk}_{\sigma'\sigma} Y_2^{i k}({\bf\hat{ r}}) \right)\hat{d}_6(r) 
+  \hat{Q}^{kl}_{\sigma'\sigma} Y_2^{kl}({\bf\hat{ r}}) \left(\delta^{ij}  \hat{e}_1(r) + Y_2^{ij}({\bf\hat{ r}}) \hat{e}_2(r)\right) \nonumber\\
&+& \hat{Q}^{ij}_{\sigma'\sigma} \hat{e}_3(r) 
+\left(\hat{Q}^{ik}_{\sigma'\sigma} Y_2^{j k}({\bf\hat{ r}}) + \hat{Q}^{jk}_{\sigma'\sigma} Y_2^{i k}({\bf\hat{ r}}) \right)\hat{e}_4(r) 
\Bigg\},
\label{t2ijdef}
\eea
where the functions $\hat{d}_i$ and $\hat{e}_i$ are given in  Appendix~\ref{dande}.
Different parametrizations of the multipole expansion of the EMT distributions have been applied in Refs.~\cite{Polyakov:2019lbq,Sun:2020wfo,Polyakov:2018rew,Panteleeva:2020ejw}. 
Using Eq.~(\ref{t2ijdef}) and the parametrization from Ref.~\cite{Panteleeva:2020ejw} we obtain
the following pressure and shear force distributions:\footnote{Notice that this interpretation in terms of the pressure and shear forces has been criticized recently in Ref.~\cite{Ji:2021mfb}.}
\bea
p_0(r) &=& N_{\phi, R,2}\int d^2\hat{n} \, \hat{d}_1(r) \,,\nonumber\\
s_0(r) &=&  N_{\phi, R,2}\int d^2\hat{n} \, \hat{d}_2(r) \,,\nonumber\\
p_2(r) &=& N_{\phi, R,2}\int d^2\hat{n} \, \frac{1}{9} \left[  -6 \hat{d}_3(r)+2
  \hat{d}_4(r)+9 \hat{d}_5(r)-6 \hat{d}_6(r)-6 \hat{e}_1(r)+2 \hat{e}_2(r)+9
   \hat{e}_3(r) -6 \hat{e}_4(r) \right] \,,\nonumber\\
s_2(r) &=& N_{\phi, R,2}\int d^2\hat{n} \, \frac{1}{6} \left[ 6 \hat{d}_3(r)-2 \hat{d}_4(r)+9
   \hat{d}_6(r) +6 \hat{e}_1(r)-2 \hat{e}_2(r)+9 \hat{e}_4(r) \right] \,,\nonumber\\
s_3(r) &=&  N_{\phi, R,2}\int d^2\hat{n} \, \frac{1}{3} \left[ -3 \hat{d}_3(r)+4 \hat{d}_4(r) -3 \hat{d}_6(r)-3 \hat{e}_1(r)+4
   \hat{e}_2(r)-3 \hat{e}_4(r) \right] \,,\nonumber\\
p_3(r) &=& N_{\phi, R,2}\int d^2\hat{n} \, \frac{1}{9} \left[ 15 \hat{d}_3(r)-2
  \hat{d}_4(r)+15 \hat{d}_6(r)+15 \hat{e}_1(r)-2 \hat{e}_2(r) +15
   \hat{e}_4(r) \right] \,.
\eea

The third part of the $ij$th components of the EMT is not conserved and it also contributes to the multipole pressure and shear force distributions 
\bea
t^{ij}_3(r)=\delta_{ij}N_{\phi, R,2}\int d^2\hat{n} \,\dfrac{d^3q}{(2\pi)^3}\Bigg\{\delta_{\sigma'\sigma}m^2\mathcal{C}_0(-{\bf q}_\perp^2)
+{\bf q}_\perp^2\hat{Q}^{kl}_{\sigma'\sigma} \hat{n}^k\hat{n}^l\mathcal{C}_1(-{\bf q}_\perp^2)+\mathcal{C}_2(-{\bf q}_\perp^2)\hat{Q}^{kl}_{\sigma'\sigma} q_\perp^kq_\perp^l
\Bigg\}
e^{-i{\bf q}\cdot{\bf r}}\,.
\eea
It can be rewritten as 
\be
t^{ij}_3(r)=\delta_{ij}\left(\delta_{\sigma'\sigma}g_1(r)+{\hat{Q}_{\sigma'\sigma}^{kl}Y^{kl}_2({\bf\hat{ r}})}  g_2(r)\right)\,,
\ee
where 
\bea
g_1(r)&=&N_{\phi,R,2}\int d^2\hat{n} \, m^2\,\tilde{\mathcal{C}}_0({\bf r}_\perp) \delta(r_\parallel)\,,\\
g_2(r)&=&-\dfrac{N_{\phi,R,2}}{2} \int d^2\hat{n} \, \left[\left( \frac{3 r_\parallel^2}{r^2}-1\right) \hat{O}_2(r_\perp)\tilde{\mathcal{C}}_1({\bf r}_\perp) \delta(r_\parallel)+\left( \frac{3 r_\perp^2}{r^2}-1\right) \hat{O}_2(r_\perp)
  \tilde{\mathcal{C}}_2({\bf r}_\perp)\delta(r_\parallel) \right] \,,
\eea
and 
\bea
\tilde{\mathcal{C}}_i(r_\perp)=\int \dfrac{d^2q_{\perp}}{(2\pi)^2}e^{-i{\bf q}_{\perp}\cdot{\bf r}_{\perp}}\mathcal{C}_i(-{\bf q}_\perp^2)\,.
\eea
In all above expressions we have dropped the corresponding subleading contributions contained in``Rest". 

It is not surprising that the normalization factors of the energy and momentum distributions diverge in the limit of sharply localized states. This is because 
for such states, the weight of the energy-momentum eigenstates with larger eigenvalues in the wave packet increases with the reduction of the localization.
On the other hand, the overall normalization of the internal pressure and shear force distributions vanish
as  these functions are related to the variation of the action with respect to the spatial metric $g_{ik}({\bf r})$. 
This variation corresponds to a change of the location of the system in three-dimensional space,
   which vanishes for  sharply localized states. 
   Notice that for spherically symmetric packets the shape of all distributions does
   not depend on the localization of the system and is uniquely determined by the corresponding form factors.

\medskip

\subsection{Gravitational densities in the Breit frame}
The ``naive'' densities in terms of the Fourier transforms of the form factors in Breit frame emerge in static approximation  by expanding 
the integrand in Eq.~(\ref{rhoint2G}) in powers of $1/m$ up to leading-order terms before performing integration. The resulting expressions have the form: 
\bea
t_\phi^{00}&=&m\int \dfrac{d^3Pd^3q}{(2\pi)^3}\left[\delta_{\sigma'\sigma}\left(A_0(-{\bf q}^2)-\dfrac{{\bf q}^2}{12M^2}{A}_1(-{\bf q}^2)\right)+\hat{Q}^{kl}_{\sigma'\sigma}q^kq^l\dfrac{{A}_1(-{\bf q}^2)}{4M^2}
\right] \phi\left({\bf P}-\frac{{\bf q}}{2}\right)\phi^\star\left({\bf P}+\frac{{\bf q}}{2}\right)e^{-i{\bf q}\cdot{\bf r}} % +O\left(m^0\right)
\,,\nonumber\\
t_\phi^{0i}&=&\int \dfrac{d^3Pd^3q}{(2\pi)^3}\left[\delta_{\sigma'\sigma}{\bf P}^i \left(A_0(-{\bf q}^2)-\dfrac{{\bf q}^2}{12M^2}{A}_1(-{\bf q}^2)\right)+{\bf P}^i\hat{Q}^{kl}_{\sigma'\sigma}q^kq^l\dfrac{{A}_1(-{\bf q}^2)}{4M^2}+\dfrac{J(-{\bf q}^2)}{2}\left(i{\bf\hat{S}}_{\sigma'\sigma}\times {\bf q}\right)^i\right]\nonumber\\
&\times&\phi\left({\bf P}-\frac{{\bf q}}{2}\right)\phi^\star\left({\bf P}+\frac{{\bf q}}{2}\right)e^{-i{\bf q}\cdot{\bf r}} % +O\left(\dfrac{1}{m}\right)
\,,\nonumber\\
t_\phi^{ij}&=&\dfrac{1}{m}\int \dfrac{d^3Pd^3q}{(2\pi)^3}\times\phi\left({\bf P}-\frac{{\bf q}}{2}\right)\phi^\star\left({\bf P}+\frac{{\bf q}}{2}\right)e^{-i{\bf q}\cdot{\bf r}}\Bigg[\dfrac{J(-{\bf q}^2)}{2}\left(P^i\left(i{\bf\hat{S}}_{\sigma'\sigma}\times {\bf q}\right)^j+P^j\left(i{\bf\hat{S}}_{\sigma'\sigma}\times {\bf q}\right)^i\right) \nonumber\\
&+&P^iP^j\left(\delta_{\sigma'\sigma}\left(A_0(-{\bf q}^2)-\dfrac{{\bf q}^2}{12M^2}{A}_1(-{\bf q}^2)\right)+\hat{Q}^{kl}_{\sigma'\sigma}q^kq^l\dfrac{{A}_1(-{\bf q}^2)}{4M^2}\right) \nonumber\\
&+&\left({\bf q}^2\delta_{ij}-q_iq_j\right)\left\{\delta_{\sigma'\sigma}\left(\dfrac{D_0(-{\bf q}^2)}{4}+\dfrac{{\bf q}^2}{48M^2}D_1(-{\bf q}^2)-\dfrac{E(-{\bf q}^2)}{3}\right)-\hat{Q}^{kl}_{\sigma'\sigma}q^kq^l\dfrac{D_1(-{\bf q}^2)}{16M^2}\right\} \nonumber\\
&+&\delta_{\sigma'\sigma}\delta_{ij}\left(\overline{f}(-{\bf q}^2)\dfrac{M^2}{6}+\overline{c}_0(-{\bf q}^2)\dfrac{M^2}{2}+\overline{c}_1(-{\bf q}^2)\dfrac{{\bf q}^2}{24}\right)-\overline{f}(-{\bf q}^2)M^2\hat{Q}^{ij}_{\sigma'\sigma}-\overline{c}_1(-{\bf q}^2)\dfrac{1}{8}\delta_{ij}\hat{Q}^{kl}_{\sigma'\sigma}q^kq^l\Bigg] \nonumber\\
&-&\dfrac{E(-{\bf q}^2)}{2}\Big(-\delta_{ij}\hat{Q}^{kl}_{\sigma'\sigma}q^kq^l+q^k(\hat{Q}^{ki}_{\sigma'\sigma}q^j+\hat{Q}^{kj}_{\sigma'\sigma}q^i)-{\bf q}^2\hat{Q}^{ij}_{\sigma'\sigma}\Big) \,.
\eea   
To consider sharply localized wave packets we expand around $R = 0$ by using the method of dimensional counting and obtain 
\bea
t_{\text{naive}}^{00}&=&m\int \dfrac{d^3q}{(2\pi)^3}\left[\delta_{\sigma'\sigma}\mathcal{E}_0^{BF}(-{\bf q}^2)+\dfrac{\mathcal{E}_2^{BF}(-{\bf q}^2)}{m^2}\hat{Q}^{kl}_{\sigma'\sigma}q^kq^l\right]
 e^{-i{\bf q}\cdot{\bf r}}\ + {\rm Rest}\,,\nonumber\\
 t_{\text{naive}}^{0i}&=&\int \dfrac{d^3q}{(2\pi)^3} \, i({\bf\hat{S}}_{\sigma'\sigma} \times{\bf q})\mathcal{J}^{BF}(-{\bf q}^2)e^{-i{\bf q}\cdot{\bf r}}\ 
 + {\rm Rest} \,,\nonumber\\
 t_{\text{naive}}^{ij}&=&\dfrac{4\pi\delta_{ij}}{3R^2m}\int d\tilde{P} \tilde{P}^4|\tilde{\phi}({\bf \tilde{P}})|^2\int \dfrac{d^3q}{(2\pi)^3}e^{-i{\bf q}\cdot{\bf r}}\left(\delta_{\sigma'\sigma}\mathcal{E}^{BF}_0(-{\bf q}^2)+\dfrac{\mathcal{E}_2^{BF}(-{\bf q}^2)}{m^2}\hat{Q}^{kl}_{\sigma'\sigma}q^kq^l\right) \nonumber\\
&+&\dfrac{1}{m}\int \dfrac{d^3q}{(2\pi)^3}e^{-i{\bf q}\cdot{\bf r}}\Bigg[\left({\bf q}^2\delta_{ij}-q_iq_j\right)\delta_{\sigma'\sigma}\mathcal{D}_0^{BF}(-{\bf q}^2)+\dfrac{\mathcal{D}_3^{BF}(-{\bf q}^2)}{m^2}
\hat{Q}^{kl}_{\sigma'\sigma}q^kq^l\left({\bf q}^2\delta_{ij}-q_iq_j\right) \nonumber\\
&+&\Big(-\delta_{ij}\hat{Q}^{kl}_{\sigma'\sigma}q^kq^l+q^k(\hat{Q}^{ki}_{\sigma'\sigma}q^j+\hat{Q}^{kj}_{\sigma'\sigma}q^i)-{\bf q}^2\hat{Q}^{ij}_{\sigma'\sigma})\Big)\mathcal{D}_2^{BF}(-{\bf q}^2) \nonumber\\
&+&\delta_{ij}\delta_{\sigma'\sigma}m^2\mathcal{C}_0^{BF}(-{\bf q}^2)-\hat{Q}^{ij}_{\sigma'\sigma}m^2\overline{f}(-{\bf q}^2)
+\delta_{ij}\hat{Q}^{kl}_{\sigma'\sigma}q^kq^l\mathcal{C}_2^{BF}(-{\bf q}^2)\Bigg]\ + {\rm Rest} \,,
\label{rhoint3NRG}
\eea 
where the explicit form of the linear combinations of the form factors, $\mathcal{E}^{BF}_i(-{\bf q}_\perp^2)$,  $\mathcal{J}^{BF}(-{\bf q}_\perp^2)$, $\mathcal{D}^{BF}_i(-{\bf q}_\perp^2)$ and $\mathcal{C}^{BF}_i(-{\bf q}_\perp^2)$ is specified in Appendix~\ref{linCofFFs}, and we have substituted $M=m$.

The $t_{\text{naive}}^{00}$, $t_{\text{naive}}^{0i}$ and the second term of $t_{\text{naive}}^{ij}$ in Eq.~(\ref{rhoint3NRG})  coincide with the corresponding expressions of spatial densities obtained as the Fourier transforms of the gravitational form factors 
 in the Breit frame in Ref.~\cite{Polyakov:2019lbq}, provided that one
takes into account the normalization factor $2m$ and performs the $1/m$ expansion up to required orders in the expressions of the last reference.

\section{Spatial densities in moving frames}
\label{movingFs}

In this section we consider a spin-1 system in the same physical state of Eq.~(\ref{statedefN2}) from the point of view of a moving frame. 
In a moving frame, our system is described by the following wave packet \cite{Hoffmann:2018edo}
\begin{eqnarray}
|\Phi, {\bf X},\sigma \rangle _{\bf v} & = & \int \frac{d^3 {p}}{\sqrt{2 E (2\pi)^3}}  \, \sqrt{\gamma \Big( 1-\frac{{\bf v} \cdot {\bf
      p}}{E} \Big)} \, \phi \big[ \Lambda_{\bf v}^{-1} {\bf p}
\big]
\, e^{-i {\bf p}\cdot{\bf X}} 
\sum_{\sigma_1}D_{\sigma_1\sigma}\Big[ W\Big( \Lambda_{\bf v}, 
    \frac{ \Lambda_{\bf v}^{-1}  {\bf p}  }{m} \Big) 
 \Big]  |p ,\sigma_1 \rangle \,,  
\label{statedefN2Moving}
\end{eqnarray}
where $\gamma = (1 - v^2)^{-1/2}$,
$E = \sqrt{m^2 + {\bf p}^2}$ and $\Lambda_{\bf v}^{-1} {\bf p} = 
     {\bf\hat { v}} \times \big(  {\bf p} \times  {\bf\hat { v}}\big)
       +  \gamma   \big( {\bf   p} \cdot  {\bf\hat  { v}}  - v E\big)  {\bf\hat { v}} $ with $\Lambda_{\bf v}$ denoting the Lorentz boost from the ZAMF to the moving frame, characterized by the vector of velocity ${\bf v}$, and ${\bf \hat{v}}={\bf v}/|{\bf v}|$.
   The $D_{\sigma_1\sigma}\left[ W\right]$ matrices in Eq.~(\ref{statedefN2Moving}) refer to the spin-1 representation of Wigner rotations
     \cite{Weinberg:1995mt}.
   
Th calculation of the local spatial densities for spin-1 systems in moving frames proceeds in close analogy to Refs.~\cite{Epelbaum:2022fjc,Panteleeva:2022khw,Panteleeva:2022uii}.
%
%\medskip
%
In the limit of sharply localized packets the leading contribution to the matrix element of the electromagnetic current in the above specified moving frame has the form:
\bea
j^\mu_{\bf v}({\bf r}) & = &\int d^3\tilde{P} \dfrac{d^3 {q}}{(2\pi)^3}\gamma\left(1-{\bf \hat{v}}\cdot{\bf\hat{\tilde{P}}}\right)\left|\tilde{\phi}\left({\bf \tilde{P}'}\right)\right|^2e^{ - i {\bf q}\cdot {\bf  r}}
D^\dagger_{\sigma'\sigma_1'}\left[ W\left( \Lambda_{\bf v},\hat {\bf m}\right) \right] D_{\sigma_1\sigma}\left[ W\left( \Lambda_{\bf v},\hat {\bf m}\right) \right]
\nonumber\\
&\times&
\hat{\tilde P}^\mu 
\Bigg\{\delta_{\sigma_1'\sigma_1}\mathcal{G}_0\left(\left( {\bf\hat{\tilde{P}}}\cdot {\bf q} \right)^2 -{\bf q}^2\right) 
+\frac{1}{2 m^2} \,\mathcal{G}_2\left(\left( {\bf\hat{\tilde{P}}}\cdot {\bf q} \right)^2 -{\bf q}^2\right)
\hat{Q}^{kl}_{\sigma_1'\sigma_1}\left(q^kq^l+\left(\hat{\tilde{\bf P}}\cdot{\bf q}\right)^2\hat{\tilde{P}}_k\hat{\tilde{P}}_l -2\left(\hat{\tilde{\bf P}}\cdot{\bf q}\right)\hat{\tilde{P}}_k q_l \right) \nonumber
\\
&+& \dfrac{1}{2m^2} \hat{Q}^{kl}_{\sigma_1'\sigma_1}\hat{\tilde{P}}_k\hat{\tilde{P}}_l 
 \left({\bf q}^2 
- \left( {\bf\hat{\tilde{P}}}\cdot {\bf q} \right)^2\right) \mathcal{G}_1\left(\left( {\bf\hat{\tilde{P}}}\cdot {\bf q} \right)^2 -{\bf q}^2\right) 
+ \dfrac{i}{m}\, \hat{\tilde{{\bf P}}}\cdot \left(\hat{{\bf S}}_{\sigma'_1\sigma_1}\times {\bf q}\right)\mathcal{M}\left(\left( {\bf\hat{\tilde{P}}}\cdot {\bf q} \right)^2 -{\bf q}^2\right)
 \Bigg\} \,, 
\label{rhoint3N}
\eea
where $\hat{\tilde P}^\mu =(1, {\bf \hat{\tilde{P}}} )$, ${\bf \hat{\tilde{P}}}={\bf {\tilde{P}}}/|{\bf {\tilde{P}}}|$,
$ {\bf\tilde P}' =
 {\bf\hat { v}} \times \big( {\bf\tilde  P} \times  {\bf\hat { v}}\big)   +
  \gamma \big({\bf\tilde   P}  \cdot  {\bf\hat  {   v}}   -   v \tilde
  P\big)    {\bf\hat { v}}$ 
and the unit vector ${\bf\hat{m}}$ is defined as ${\bf\hat{m}}\equiv {\bf\hat {\tilde P}}'$.
The combinations of form factors in Eq.~(\ref{rhoint3N}) are defined as in the ZAMF, i.e. by
Eq.~(\ref{ourGs}).
We change the integration variable ${\bf  \tilde P} \to {\bf \tilde P}'$ and define a vector-valued function
\begin{equation}
  {\bf n} \big({\bf v},  { {\bf \hat m}} \big) = { {\bf \hat v}} \times \big( {\bf  \hat m} \times  { {\bf \hat v}}\big)
+ \gamma \big( {\bf  \hat m} \cdot  { {\bf \hat v}}  + v )  {{\bf \hat v}} \,.
\end{equation}
Given that ${\bf \tilde P} =
 {{\bf \hat  v}} \times \big( {\bf \tilde  P}' \times  {{\bf \hat v}}\big)   +
  \gamma \big( {\bf \tilde  P}'  \cdot  { {\bf  \hat  v}}   +   v \tilde
  P'\big)    {{\bf \hat v}}$, it follows that  $ { {\bf \hat n}} = {\bf  {\hat {\tilde P}}} $.  
The Jacobian of the change of variables ${\bf \tilde P} \to {\bf\tilde P}'$
cancels the first factor in the integrands and after some simplifications we obtain
\bea
j^\mu_{\bf v}({\bf r}) & = & \frac{1}{4\pi} \int d \hat{\bf m} \, \dfrac{d^3 {q}}{(2\pi)^3} 
\, e^{ - i {\bf q}\cdot {\bf  r}}
D^\dagger_{\sigma'\sigma_1'}\left[ W\left( \Lambda_{\bf v},\hat {\bf m}\right) \right] D_{\sigma_1\sigma}\left[ W\left( \Lambda_{\bf v},\hat {\bf m}\right) \right] \hat n^\mu\Bigg\{ 
\frac{i }{m} \hat{\bf n}\cdot \left( {\bf \hat{S}}_{\sigma_1'\sigma_1} \times {\bf q} \right)\mathcal{M}\left(-{\bf q}_\perp^2 \right)  \nonumber\\
&+&  \delta_{\sigma_1'\sigma_1} \mathcal{G}_0\left( -{\bf q}_\perp^2 \right) 
+
\frac{{\bf q}_\perp^2}{2m^2} \hat{Q}^{kl}_{\sigma_1' \sigma_1} \hat{n}^k \hat{ n}^l 
\mathcal{G}_1\left( -{\bf q}_\perp^2\right)  
+
 \hat{Q}^{kl}_{\sigma_1' \sigma_1} \dfrac{ q^k_\perp q^l_\perp}{2 m^2} 
\mathcal{G}_2\left(-{\bf q}_\perp^2\right) \Bigg\} \, , 
\label{rhoint3NR2}
\eea
where $\hat n^\mu=(1,\hat{\bf n})$ and ${\bf q}_\perp^2= {\bf q}^2  -\left(\hat{\bf n} \cdot {\bf q} \right)^2 $.

\medskip

In the IMF with $v\to 1$ and $\gamma\to\infty$, $\hat{\bf n}$ turns to $\hat{\bf v}$ and using explicit form of the Wigner rotation 
matrices, and  the integration over $\hat{\bf m}$ 
can be carried out explicitly. The resulting expression has the form: 
\bea
j^\mu_{\bf v}({\bf r}) & = & \int \dfrac{d^3 {q}}{(2\pi)^3} 
\, e^{ - i {\bf q}\cdot {\bf  r}} \, 
\hat v^\mu\Bigg\{ \delta_{\sigma'\sigma}\,  {\cal G}_0\left( -{\bf q}_\perp^2 \right)
+ \frac{i }{2m}  \hat{\bf v} \cdot\left( {\bf \hat{S}}_{\sigma'\sigma} \times {\bf q} \right)   \, 
{\cal M} \left( -{\bf q}_\perp^2 \right) 
\nonumber\\ & + & 
 \frac{1}{6 m^2}\left( q^k_\perp q^l_\perp  +
\frac{{\bf q}_\perp^2}{2} \, \hat{v}^k\hat{v}^l \right) \hat{Q}^{kl}_{\sigma' \sigma} {\cal G}_2\left(  -{\bf q}_\perp^2 \right) \Bigg\} \, .
\label{rhoint3NR3x}
\eea

\medskip

Analogously to the electromagnetic current, the matrix element of the EMT in a moving frame for a sharply localized state can be written as
\bea
t_{\phi}^{00}&=&\int d{\bf \hat{m}}\dfrac{d\tilde{P}'\tilde{P}'^2d^3q}{(2\pi)^3}\left|\tilde{\phi}\left(\tilde{P}'\right)\right|e^{-i{\bf q}\cdot{\bf r}}D^\dagger_{\sigma'\sigma_1'}\left[ W\left( \Lambda_{\bf v},\hat {\bf m}\right) \right] D_{\sigma_1\sigma}\left[ W\left( \Lambda_{\bf v},\hat {\bf m}\right) \right]\dfrac{\gamma( \tilde P'+v \tilde P'_\parallel  )}{R}
\nonumber\\
&\times&
\Bigg\{\delta_{\sigma_1'\sigma_1}\mathcal{E}_0(-{\bf q}_\perp^2)+\hat{Q}_{\sigma_1'\sigma_1}^{kl}\hat{n}^k\hat{n}^l \dfrac{{\bf q}_\perp^2}{m^2}\mathcal{E}_1(-{\bf q}_\perp^2)+\dfrac{\mathcal{E}_2(-{\bf q}_\perp^2)}{m^2}\hat{Q}_{\sigma_1'\sigma_1}^{kl}q_\perp^k q^l_\perp+{\bf \hat n}\cdot({\bf\hat{S}}_{\sigma_1'\sigma_1}\times   {\bf q})\dfrac{i\mathcal{J}(-{\bf q}_\perp^2)}{m}
\Bigg\} \ +{\rm Rest} \,,\nonumber\\
t_{\phi}^{0i}&=&\int d{\bf \hat{m}}\dfrac{d\tilde{P}'\tilde{P}'^2d^3q}{(2\pi)^3}\left|\tilde{\phi}\left(\tilde{P}'\right)\right|e^{-i{\bf q}\cdot{\bf r}}D^\dagger_{\sigma'\sigma_1'}\left[ W\left( \Lambda_{\bf v},\hat {\bf m}\right) \right] D_{\sigma_1\sigma}\left[ W\left( \Lambda_{\bf v},\hat {\bf m}\right) \right]\dfrac{\gamma( \tilde P'+v \tilde P'_\parallel  )}{R}
\nonumber\\
&\times&{\bf \hat{n}}
\Bigg\{\dfrac{i\mathcal{J}(-{\bf q}_\perp^2)}{m}
({\bf\hat{S}}_{\sigma_1'\sigma_1}\times   {\bf q})\cdot{\bf \hat n}+\delta_{\sigma_1'\sigma_1}\mathcal{E}_0(-{\bf q}_\perp^2)+\hat{Q}_{\sigma_1'\sigma_1}^{kl}\hat{n}^k\hat{n}^l \dfrac{{\bf q}_\perp^2}{m^2}\mathcal{E}_1(-{\bf q}_\perp^2)+\dfrac{\mathcal{E}_2(-{\bf q}_\perp^2)}{M^2}\hat{Q}_{\sigma_1'\sigma_1}^{kl}q_\perp^k q^l_\perp\Bigg\}\  +{\rm Rest}\,,\nonumber\\
t_{\phi}^{ij}&=&\int d{\bf \hat{m}}\dfrac{d\tilde{P}'\tilde{P}'^2d^3q}{(2\pi)^3}\left|\tilde{\phi}\left(\tilde{P}'\right)\right|e^{-i{\bf q}\cdot{\bf r}}D^\dagger_{\sigma'\sigma_1'}\left[ W\left( \Lambda_{\bf v},\hat {\bf m}\right) \right] D_{\sigma_1\sigma}\left[ W\left( \Lambda_{\bf v},\hat {\bf m}\right) \right]\dfrac{\gamma( \tilde P'+v \tilde P'_\parallel  )}{R}
\nonumber\\
&\times&\hat{n}^i\hat{n}^j\Bigg\{\delta_{\sigma_1'\sigma_1}\mathcal{E}_0(-{\bf q}_\perp^2)+\hat{Q}_{\sigma_1'\sigma_1}^{kl}\hat{n}^k\hat{n}^l\dfrac{{\bf q}^2_\perp}{m^2}\mathcal{E}_1(-{\bf q}_\perp^2)+\dfrac{\mathcal{E}_2(-{\bf q}_\perp^2)}{m^2}\hat{Q}_{\sigma_1'\sigma_1}^{kl}q_\perp^kq_\perp^l+\dfrac{i\mathcal{J}(-{\bf q}_\perp^2)}{m}
({\bf\hat{S}}_{\sigma_1'\sigma_1}\times   {\bf q})\cdot{\bf \hat n}\Bigg\} \nonumber\\
&+&
\int d{\bf \hat{m}}\dfrac{d\tilde{P}'\tilde{P}'^2d^3q}{(2\pi)^3}\left|\tilde{\phi}\left(\tilde{P}'\right)\right|e^{-i{\bf q}\cdot{\bf r}}D^\dagger_{\sigma'\sigma_1'}\left[ W\left( \Lambda_{\bf v},\hat {\bf m}\right) \right] D_{\sigma_1\sigma}\left[ W\left( \Lambda_{\bf v},\hat {\bf m}\right) \right] \dfrac{R}{2\gamma( \tilde P'+v \tilde P'_\parallel  )}
\nonumber\\
&\times&\Bigg\{\left(\delta_{ij}{\bf q}^2_{\perp}-q_iq_j\right)\Bigg[\delta_{\sigma_1'\sigma_1}\mathcal{D}_0(-{\bf q}_\perp^2)+\dfrac{\mathcal{D}_2(-{\bf q}_\perp^2)}{m^2}\hat{Q}_{\sigma_1'\sigma_1}^{kl}q_\perp^kq_\perp^l+\left( \hat{Q}_{\sigma_1'\sigma_1}^{kl}\hat{n}^k\hat{n}^l\dfrac{{\bf q}^2_\perp}{m^2}- 2 \dfrac{i}{m}{\bf \hat n}\cdot({\bf\hat{S}}_{\sigma_1'\sigma_1}\times   {\bf q})\right)\mathcal{D}_1(-{\bf q}_\perp^2)\Bigg]\nonumber\\
&+&\delta_{ij}\Bigg[\delta_{\sigma_1'\sigma_1}m^2\mathcal{C}_0(-{\bf q}_\perp^2)+\mathcal{C}_1(-{\bf q}_\perp^2)\left({\bf q}_\perp^2\hat{Q}_{\sigma_1'\sigma_1}^{kl}\hat{n}^k\hat{n}^l - 2\, \dfrac{i}{m}({\bf\hat{S}}_{\sigma_1'\sigma_1}\times   {\bf q})\cdot{\bf \hat n}
\right)+\mathcal{C}_2(-{\bf q}_\perp^2)\hat{Q}_{\sigma_1'\sigma_1}^{kl}q_\perp^kq_\perp^l\Bigg]
\Bigg\}+{\rm Rest}\,,
\eea
where in exact analogy to the ZAMF we show explicitly only the leading-order contributions for each form factor, and the explicit form of the linear combinations of the form factors, $\mathcal{E}_i(-{\bf q}_\perp^2)$,  $\mathcal{J}(-{\bf q}_\perp^2)$, $\mathcal{D}_i(-{\bf q}_\perp^2)$ and $\mathcal{C}_i(-{\bf q}_\perp^2)$ is specified in Appendix~\ref{linCofFFs} .

In the IMF with $\hat {\bf n} \stackrel{v \to 1}{\longrightarrow} \hat {\bf v}$ and $\gamma\to\infty$, the integration over $\hat{\bf m}$ 
can be carried out explicitly using the explicit form of the Wigner rotation matrices. The integration over $\hat{\bf m}$ is done in full analogy to the electromagnetic case. The resulting expressions after dropping the ``Rest" contributions have the form:   
\bea
t_{\phi}^{00}&=&4\pi\gamma N_{\phi,R}\int \dfrac{d^3q}{(2\pi)^3}e^{-i{\bf q}\cdot{\bf r}}
\Bigg\{\delta_{\sigma'\sigma}\mathcal{E}_0(-{\bf q}_\perp^2) 
+ \left( q_\perp^k q^l_\perp +\dfrac{{\bf q}_\perp^2}{2} \, \hat{v}^k\hat{v}^l \right) \hat{Q}^{kl}_{\sigma'\sigma}  \dfrac{\mathcal{E}_2 (-{\bf q}_\perp^2)}{2m^2}  
+{\bf \hat v}\cdot({\bf\hat{S}}_{\sigma'\sigma}\times   {\bf q})\dfrac{2i\mathcal{J}(-{\bf q}_\perp^2)}{3m}
\Bigg\}\nonumber\\
t_{\phi}^{0i}&=&4\pi \gamma N_{\phi,R}\int\dfrac{d^3 q}{(2\pi)^3}e^{-i{\bf q}\cdot{\bf r}}
{\bf \hat{v}}
\Bigg\{\delta_{\sigma'\sigma}\mathcal{E}_0(-{\bf q}_\perp^2) 
+ \left( q_\perp^k q^l_\perp +\dfrac{{\bf q}_\perp^2}{2} \, \hat{v}^k\hat{v}^l \right) \hat{Q}^{kl}_{\sigma'\sigma}  \dfrac{\mathcal{E}_2 (-{\bf q}_\perp^2)}{2m^2} 
%\dfrac{\mathcal{E}_2(-{\bf q}_\perp^2)}{2M^2}  \left(\hat{Q}_{\sigma'\sigma}^{kl}
%\,q_\perp^k q^l_\perp{\color{red}+\dfrac{{\bf q}_\perp^2}{2}\hat{v}^k\hat{v}^l\hat{Q}^{kl}_{\sigma'\sigma}}\right)  
+
{\bf \hat v}\cdot({\bf\hat{S}}_{\sigma'\sigma}\times   {\bf q})\dfrac{2 i\mathcal{J}(-{\bf q}_\perp^2)}{3m}\Bigg\},\nonumber\\
t_{\phi}^{ij}&=&4\pi \gamma N_{\phi,R}\int \dfrac{d^3q}{(2\pi)^3}e^{-i{\bf q}\cdot{\bf r}}
\hat{v}^i\hat{v}^j\Bigg\{\delta_{\sigma'\sigma}\mathcal{E}_0(-{\bf q}_\perp^2)
+ \left( q_\perp^k q^l_\perp +\dfrac{{\bf q}_\perp^2}{2} \, \hat{v}^k\hat{v}^l \right) \hat{Q}^{kl}_{\sigma'\sigma}  \dfrac{\mathcal{E}_2 (-{\bf q}_\perp^2)}{2m^2} 
% + \dfrac{\mathcal{E}_2(-{\bf q}_\perp^2)}{2M^2} \left(\hat{Q}_{\sigma'\sigma}^{kl}
% \, q_\perp^k q^l_\perp {\color{red}+\dfrac{{\bf q}_\perp^2}{2}\hat{v}^k\hat{v}^l\hat{Q}^{kl}_{\sigma'\sigma}}\right) 
+{\bf \hat v}\cdot({\bf\hat{S}}_{\sigma'\sigma}\times   {\bf q})\dfrac{2i\mathcal{J}(-{\bf q}_\perp^2)}{3m}
\Bigg\}\nonumber\\
&-&
2\pi N_{\phi,R,2}\dfrac{\ln(1-v)}{\gamma}\int \dfrac{d^3q}{(2\pi)^3}e^{-i{\bf q}\cdot{\bf r}}\Bigg\{\left(\delta_{ij}{\bf q}^2_{\perp}-q_iq_j\right)\bigg[
 \delta_{\sigma'\sigma}\mathcal{D}_0(-{\bf q}_\perp^2)  + \hat{v}^k\hat{v}^l\hat{Q}^{kl}_{\sigma'\sigma} \,
 \frac{{\bf q}_\perp^2}{m^2} \left( \mathcal{D}_1(-{\bf q}_\perp^2)  -  \dfrac{\mathcal{D}_2(-{\bf q}_\perp^2)}{2}\right)  \bigg] \nonumber\\
&+&\delta_{ij}\bigg[\delta_{\sigma'\sigma}m^2\mathcal{C}_0(-{\bf q}_\perp^2)+ \hat{v}^k\hat{v}^l\hat{Q}^{kl}_{\sigma'\sigma} \, 
{\bf q}_\perp^2 \left( \mathcal{C}_1(-{\bf q}_\perp^2)  -  \dfrac{\mathcal{C}_2(-{\bf q}_\perp^2)}{2}\right)  \bigg]
\Bigg\}\,. 
\label{GDMF}
\eea
The expressions in Eq.~(\ref{GDMF}) have the same interpretation as their analogues in the ZAMF.

Comparing Eqs.~(\ref{rhoint3NR3x}) and (\ref{rhoint3Na}),
%as well as Eqs.~(\ref{GDMF}) and (\ref{GSDZ}) 
it is easily seen that by integrating the IMF expressions over all directions of $ {\bf \hat v}$, one reconstructs the three-dimensional ZAMF electric charge density (the symmetric part of $j^0({\bf r})$) exactly and the three-dimensional magnetic charge density with the additional factor of $1/2$, the same holds true for systems with spin-0 and spin-$1/2$ studied in Refs.~\cite{Epelbaum:2022fjc,Panteleeva:2022khw}. The more complicated quadrupole charge density  in the ZAMF can not be reproduced from the moving IMF expressions. 
%This differs from the case of the charge density of spin-0 systems considered in Ref.~\cite{Epelbaum:2022fjc}.
This is because the Lorentz boosts to moving frames amount not only to Lorentz contractions but also involve Wigner rotations, which  modify the non-symmetric quadrupole structure. 
Comparing  Eqs.~(\ref{GDMF}) and (\ref{GSDZ}) one sees again that the densities with spin structures $\sim \delta_{\sigma'\sigma}$ and $\sim {\bf S}_{\sigma'\sigma}$ in the ZAMF can be restored by averaging the IMF expressions up the normalization factor, while the quadrupole structure $\sim\hat{Q}^{kl}_{\sigma'\sigma}$ can not be obtained this way. 
   
\section{Summary}
\label{summary}

In this work we considered matrix elements of the electromagnetic current and the EMT operators for spin-1 systems calculated for sharply localized 
one-particle states. We obtained the resulting expressions of the  local spatial distributions in terms of  the form factors in the ZAMF as well as in moving frames. By considering the static approximation we also obtained the traditional 
expressions in terms of the form factors in the Breit frame. Next we discussed the physical interpretation of obtained spatial densities.
Having calculated the spatial densities in the IMF, we found that the expressions for the ZAMF densities coincide with the ones obtained by integrating the corresponding IMF expressions over all possible directions, as was also found for spin-0 and spin-1/2 systems. The only exceptions are the quadrupole densities for spin-1 systems, where the mismatch can be traced back to the fact that Wigner rotations modify the quadrupole structure.

As the next step we plan to apply the obtained results to the electromagnetic and gravitational densities of the deuteron within the framework of the low-energy effective field theory of QCD.

\acknowledgements
%{\it Acknowledgements:}
This work was supported in part by BMBF (Grant No. 05P21PCFP1), by
DFG and NSFC through funds provided to the Sino-German CRC 110
``Symmetries and the Emergence of Structure in QCD'' (NSFC Grant
No. 11621131001, DFG Project-ID 196253076 - TRR 110),
by ERC  NuclearTheory (grant No. 885150),  %and ERC EXOTIC (grant No. 101018170),
by CAS through a President's International Fellowship Initiative (PIFI)
(Grant No. 2018DM0034), by the VolkswagenStiftung
(Grant No. 93562), by the EU Horizon 2020 research and
innovation programme (STRONG-2020, grant agreement No. 824093),
and by the MKW NRW
under the funding code NW21-024-A.
%by the Heisenberg-Landau Program 2021. 

\appendix

\section{Distinguishing between $m$ and $M$.}
\label{dist}

Below we demonstrate the importance of distinguishing between $m$ and $M$ when taking the static limit.
To obtain the charge density in the static approximation we expand 
the integrand in Eq.~(\ref{rhoint2N}) in powers of $1/m$ and keep only the leading order term. Then we expand the integrand in powers of $R$ around $R = 0$ and keep terms up to the zeroth order. 
Integration over $P$ now results in 
\bea
j^0_{\text{naive}}({\bf r})&=&\int \dfrac{d^3 q}{(2\pi)^3}e^{-i{\bf q}\cdot {\bf r}}\Bigg\{\delta_{\sigma\sigma'}\left(G_1(-{\bf q}^2)+\dfrac{{\bf q}^2}{6M^2}G_3(-{\bf q}^2)\right)-G_3(-{\bf q}^2)\hat{Q}^{km}\dfrac{q^kq^m}{2M^2} 
\Bigg\}\,.
%=\nonumber\\
%&=&\int \dfrac{d^3 q}{(2\pi)^3}e^{-i{\bf q}\cdot {\bf r}}\left(G_C(-{\bf q}^2)\delta_{\sigma\sigma'}+\dfrac{G_Q(-{\bf q}^2)}{2M^2}\hat{Q}^{km}q^mq^k\right)\equiv\nonumber\\
%&\equiv& \delta_{\sigma\sigma'}\rho_C^{naive}(r)+\hat{Q}^{km}Y^{km}_2({ r})\rho_Q^{naive}(r),
\label{chdstapprox}
\eea
By substituting $M=m$ in Eq.~(\ref{chdstapprox}) we obtain the expression displayed in Eq.~(\ref{staticEM}). 
Notice that there would be no contribution of the form factor $G_3$ in Eq.~(\ref{chdstapprox}) if we would not distinguish between $m$ and $M$ and keep only the leading order term of the $1/m$ expansion. One might think that the expression of the charge density given in Eq.~(\ref{staticEM}) could be also obtained by taking $M=m$ from the very beginning and keeping the terms up to $1/m^2$ in the $1/m$ expansion of the integrand. Doing so we obtain 
\bea
j^0_{\text{naive}}({\bf r})\!\!&=&\!\! \int\!\! \dfrac{\tilde P^2 d \tilde P d^2\hat{n}  d^3 q}{(2\pi)^3} \, \tilde\phi\left( |\tilde {\bf P} | \right)  \tilde\phi^\star\left( |\tilde {\bf P} | \right)e^{-i{\bf q}\cdot {\bf r}} \Bigg\{\delta_{\sigma\sigma'}\left(G_1(-{\bf q}^2)+\dfrac{{\bf q}^2}{6m^2}G_3(-{\bf q}^2)\right)-G_3(-{\bf q}^2)\hat{Q}^{km}\dfrac{q^kq^m}{2m^2} \nonumber\\
\!\!&+& \!\!
\frac{\delta_{\sigma'\sigma}}{6m^2R^2}  \left( 6 \tilde P^2 {\bf q}_\parallel^2 G_1'(-{\bf q}^2) + {\bf q}^2 R^2 G_1(-{\bf q}^2) 
-{\bf q}^2 R^2 G_2(-{\bf q}^2)\right) 
\! +\! \hat{Q}^{kl}_{\sigma'\sigma}\dfrac{q^kq^l}{2m^2}\left({G}_2(-{\bf q}^2)-G_1(-{\bf q}^2)\right)\!\!
\Bigg\}\,.
\label{chdstapprox2}
\eea
Eq.~(\ref{chdstapprox2}) apparently does not reproduce the expression of Eq.~(\ref{staticEM}). Moreover, it contains terms which diverge in $R \to 0$ limit. 
This is caused by the non-commutativity of the $1/m$ expansion with the expansion around $R=0$.

\section{Spin operators}
\label{append}
The spin ($S$) and quadrupole ($Q$) operators defined in terms of the polarization vectors of Eq.(\ref{PolarVect}) (for more details  see Ref.~\cite{Varshalovich}):   
\bea
\langle \sigma'|\hat{S}^i|\sigma\rangle &\equiv&(\hat{S}^i)_{\sigma'\sigma}=-i\epsilon^{ijk}\epsilon^{\star j}_{\sigma'}\epsilon^k_{\sigma}, \nonumber\\
\hat{Q}^{ij}_{\sigma'\sigma} &=& \dfrac{1}{2}\left(\hat{S}^i\hat{S}^j+\hat{S}^j\hat{S}^i-\dfrac{2}{3}S(S+1)\delta^{ij}\right)_{\sigma'\sigma} = \dfrac{1}{3}\delta^{ij}\delta_{\sigma\sigma'} 
- \dfrac{1}{2}\left(\hat{\epsilon}^{\star i}_{\sigma'}\hat{\epsilon}^{j}_{\sigma}+\hat{\epsilon}^{\star j}_{\sigma'}\hat{\epsilon}^{i}_{\sigma}\right).
\eea 
Using these definitions the following useful relations can be obtained:
\bea
{\boldsymbol\epsilon}_\sigma({\boldsymbol{\hat{\epsilon}}}_{\sigma'}^\star\cdot {\bf q})-{\boldsymbol\epsilon}^\star_{\sigma'}({\boldsymbol{\hat{\epsilon}}}_\sigma\cdot {\bf q})&=&i({\bf\hat{S}}_{\sigma'\sigma}\times{\bf q})     ,\\
({\boldsymbol{\hat{\epsilon}}}_\sigma\cdot {\bf q})({\boldsymbol{\hat{\epsilon}}}^{\star}_{\sigma'}\cdot {\bf q})&=& \dfrac{{\bf q}^2}{3}\delta_{\sigma\sigma'}-\hat{Q}^{kl}_{\sigma'\sigma} q^kq^l,\\
({\boldsymbol{\hat{\epsilon}}}_\sigma\cdot {\bf \hat n})({\boldsymbol{\hat{\epsilon}}}^{\star}_{\sigma'}\cdot {\bf\hat n})&=& \dfrac{1}{3}\delta_{\sigma\sigma'} 
-\hat{Q}^{kl}_{\sigma'\sigma}\hat{n}^k\hat{n}^l,\\
({\boldsymbol{\hat{\epsilon}}}_\sigma\cdot {\bf \hat n})({\boldsymbol{\hat{\epsilon}}}^{\star}_{\sigma'}\cdot {\bf q})+({\boldsymbol{\hat{\epsilon}}}_\sigma\cdot {\bf q})({\boldsymbol{\hat{\epsilon}}}^{\star}_{\sigma'}\cdot {\bf \hat n})&=&  \dfrac{2}{3}({\bf \hat{n}}\cdot{\bf q})\delta_{\sigma\sigma'}-2\hat{Q}^{nk}_{\sigma'\sigma} \hat{n}^kq^l,\\
{\boldsymbol{\hat{\epsilon}}}_\sigma^i {\boldsymbol{\hat{\epsilon}}}^{\star\,j}_{\sigma'} + {\boldsymbol{\hat{\epsilon}}}_\sigma^j {\boldsymbol{\hat{\epsilon}}}^{\star\,i}_{\sigma'} &=&  \dfrac{2}{3}\, \delta^{ij}\delta_{\sigma\sigma'}-2\hat{Q}^{ij}_{\sigma'\sigma}\,,\\
({\boldsymbol{\hat{\epsilon}}}^\star_{\sigma'}\times{\boldsymbol{\hat{\epsilon}}}_\sigma)&=& i{\bf\hat{S}}_{\sigma'\sigma}.
\eea

Multipole tensor of the $n$th rank is defined for $r\neq 0$ as \cite{Polyakov:2018rew}:
\be
Y^{i_1,i_2,...,i_n}_n({\bf\hat{ r}})=\dfrac{(-1)^n}{(2n-1)!!}r^{n+1}\partial^{i_1}...\partial^{i_n}\dfrac{1}{r}\,.
\label{MultipoleTensDef}
\ee
From Eq.~(\ref{MultipoleTensDef}) follows in particular:
\be
Y_0({\bf\hat{ r}})=1,\ \ Y^i_1({\bf\hat{ r}})=\dfrac{r^i}{r},\ \ Y^{ij}_2({\bf\hat{ r}})=\dfrac{r^ir^j}{r^2}-\dfrac{1}{3}\delta^{ij}\,.
\ee

\section{Linear combinations of the gravitational form factors}
\label{linCofFFs}
Linear combinations of gravitational form factors in the ZAMF:
\bea
\mathcal{E}_0(-{\bf q}_\perp^2)&=&A_0(-{\bf q}_\perp^2)-\dfrac{{\bf q}_\perp^2}{12 m^2}A_1(-{\bf q}_\perp^2)+\nonumber\\
&+&\dfrac{{\bf q}_\perp^2}{12m^2}\left(4J(-{\bf q}_\perp^2)-2E(-{\bf q}_\perp^2)-2A_0(-{\bf q}_\perp^2)+A_1(-{\bf q}_\perp^2)\dfrac{{\bf q}_\perp^2}{4 m^2}\right)+\dfrac{M^2}{3m^2}\overline{f}(-{\bf q}_\perp^2)\,,\nonumber\\
\mathcal{E}_2(-{\bf q}_\perp^2)&=&\frac{A_1(-{\bf q}_\perp^2)}{4} \,,\nonumber\\
\mathcal{E}_1(-{\bf q}_\perp^2)&=&\dfrac{1}{2}\left(A_0(-{\bf q}_\perp^2)+E(-{\bf q}_\perp^2)-2J(-{\bf q}_\perp^2)-A_1(-{\bf q}_\perp^2)\dfrac{{\bf q}_\perp^2}{8 m^2}\right)-\dfrac{M^2}{{\bf q}_\perp^2}\overline{f}(-{\bf q}_\perp^2)\,,\nonumber\\
\mathcal{J}(-{\bf q}_\perp^2)&=&J(-{\bf q}_\perp^2)-A_0(-{\bf q}_\perp^2)+A_1(-{\bf q}_\perp^2)\dfrac{{\bf q}_\perp^2}{8 m^2}\,,\nonumber\\
\mathcal{D}_0(-{\bf q}_\perp^2)&=&\dfrac{D_0(-{\bf q}_\perp^2)}{2}+\dfrac{{\bf q}^2_{\perp}}{24 m^2}D_1(-{\bf q}_\perp^2)-\dfrac{{\bf q}^2_{\perp}}{12m^2}\left(D_0(-{\bf q}_\perp^2)+\dfrac{{\bf q}_\perp^2}{8 m^2}D_1(-{\bf q}_\perp^2)\right)\,, \nonumber\\
\mathcal{D}_1(-{\bf q}_\perp^2)&=&\dfrac{1}{4}\Bigg[D_0(-{\bf q}_\perp^2)+D_1(-{\bf q}_\perp^2)\dfrac{{\bf q}_\perp^2}{8 m^2}\Bigg]\,, \nonumber\\
\mathcal{D}_2(-{\bf q}_\perp^2)&=&-\dfrac{1}{8}D_1(-{\bf q}_\perp^2)\,,\nonumber\\
\mathcal{C}_0(-{\bf q}_\perp^2)&=&\overline{c}_0(-{\bf q}_\perp^2)+\dfrac{{\bf q}_\perp^2}{12 m^2}\overline{c}_1(-{\bf q}_\perp^2)-\dfrac{{\bf q}_\perp^2}{6m^2}\left(\overline{c}_0(-{\bf q}_\perp^2)+\dfrac{{\bf q}_\perp^2}{8 m^2}\overline{c}_1(-{\bf q}_\perp^2)\right)\,,\nonumber\\
\mathcal{C}_1(-{\bf q}_\perp^2)&=&\dfrac{1}{2}\left(\overline{c}_0(-{\bf q}_\perp^2)+\dfrac{{\bf q}_\perp^2}{8 m^2}\overline{c}_1(-{\bf q}_\perp^2)\right),\nonumber\\
\mathcal{C}_2(-{\bf q}_\perp^2)&=&-\dfrac{\overline{c}_1(-{\bf q}_\perp^2)}{4}\,.
\eea

Linear combinations of gravitational form factors in the Breit frame:
\bea
\mathcal{E}_0^{BF}(-{\bf q}^2)&=&A_0(-{\bf q}^2)-\dfrac{{\bf q}^2}{12 m^2}A_1(-{\bf q}^2) % +O\left(\dfrac{1}{m}\right)
\,,\nonumber\\
\mathcal{E}_2^{BF}(-{\bf q}^2)&=&\dfrac{A_1(-{\bf q}^2)}{4} %+O\left(\dfrac{1}{m}\right)
\,,\nonumber\\
\mathcal{J}^{BF}(-{\bf q}^2)&=&\dfrac{J(-{\bf q}^2)}{2} %+O\left(\dfrac{1}{m}\right)
\,,\nonumber\\
\mathcal{D}_0^{BF}(-{\bf q}^2)&=&\dfrac{D_0(-{\bf q}^2)}{4}+\dfrac{{\bf q}^2}{48 m^2}D_1(-{\bf q}^2)-\dfrac{E(-{\bf q}^2)}{3} %+O\left(\dfrac{1}{m}\right)
\,,\nonumber\\
\mathcal{D}_2^{BF}(-{\bf q}^2)&=&-\dfrac{E(-{\bf q}^2)}{2} % +O\left(\dfrac{1}{m}\right)
\,,\nonumber\\
\mathcal{D}_3^{BF}(-{\bf q}^2)&=&-\dfrac{D_1(-{\bf q}^2)}{16}% +O\left(\dfrac{1}{m}\right)
\,,\nonumber\\
\mathcal{C}_0^{BF}(-{\bf q}^2)&=&\overline{f}(-{\bf q}^2)\dfrac{1}{12}+\overline{c}_0(-{\bf q}^2)\dfrac{1}{2}+\dfrac{{\bf q}^2}{24 m^2}\overline{c}_1(-{\bf q}^2) %+O\left(\dfrac{1}{m}\right)
\,,\nonumber\\
\mathcal{C}_2^{BF}(-{\bf q}^2)&=&-\dfrac{\overline{c}_1(-{\bf q}^2)}{8} % +O\left(\dfrac{1}{m}\right)
\,.
\eea

\section{The coefficients $\hat d_i$ and $\hat e_i$}
\label{dande}
The differential operators $\hat{d}_i$ and $\hat{e}_i$:
\bea
\hat{d}_1(r)&=& \left(\dfrac{1}{3}\hat{O}_2(r_\parallel)- \frac{2}{3}\hat{O}_2(r_\perp)\right)\tilde{\mathcal{D}}_0(r_\perp) \delta(r_\parallel) \,,\nonumber\\
\hat{d}_2(r)&=& \Biggl[  -\frac{1}{2} \hat{O}_2(r_\perp) -\dfrac{1}{2} \hat{O}_2(r_\parallel)
+\frac{3}{r^2} r_\perp^k r_\parallel^l\dfrac{d^2}{dr^k_\perp dr^l_\parallel}
+\frac{3}{2r^2} \left( r_\perp^k r_\perp^l\dfrac{d^2}{dr^k_\perp dr^l_\perp} +r_\parallel^k r_\parallel^l\dfrac{d^2}{dr^k_\parallel dr^l_\parallel} \right)  \Biggr]\tilde{\mathcal{D}}_0(r_\perp) \delta(r_\parallel)
\,,\nonumber\\
\hat{d}_3(r)&=&\left[\frac{1}{2}\left( 3\,\frac{r_\parallel^2}{r^2}-1\right)  \hat{O}_2(r_\perp)+\hat{O}_3(r_\perp,r_\parallel)\right]\hat{O}_2(r_\perp) \dfrac{\tilde{\mathcal{D}}_1(r_\perp)}{m^2} \delta(r_\parallel) \,,\nonumber\\
\hat{d}_4(r)&=& \hat{O}_4(r_\perp, r_\parallel)   \hat{O}_2(r_\perp) \dfrac{\tilde{\mathcal{D}}_1(r_\perp)}{m^2} \delta(r_\parallel)  \,,\nonumber\\
\hat{d}_5(r)&=& \hat{O}_5(r_\perp,r_\parallel)  \hat{O}_2(r_\perp) \dfrac{\tilde{\mathcal{D}}_1(r_\perp)}{m^2} \delta(r_\parallel) \,,\nonumber\\
\hat{d}_6(r)&=&\hat{O}_6(r_\perp,r_\parallel) \hat{O}_2(r_\perp)   \dfrac{\tilde{\mathcal{D}}_1(r_\perp)}{m^2} \delta(r_\parallel) \, ,\nonumber\\
\hat{e}_1(r)&=&\left[ \frac{1}{2}\left( 3\,\frac{r_\perp^2}{r^2}-1\right)  \hat{O}_2(r_\perp)+\hat{O}_3(r_\parallel, r_\perp) \right] \hat{O}_2(r_\perp) \dfrac{\tilde{\mathcal{D}}_2(r_\perp)}{m^2} \delta(r_\parallel) \, ,\nonumber\\
\hat{e}_2(r)&=& \hat{O}_4(r_\parallel,r_\perp) \hat{O}_2(r_\perp) \dfrac{\tilde{\mathcal{D}}_2(r_\perp)}{m^2} \delta(r_\parallel)  \,, \nonumber\\
\hat{e}_3(r)&=& \hat{O}_5(r_\parallel,r_\perp) \hat{O}_2(r_\perp)  \dfrac{\tilde{\mathcal{D}}_2(r_\perp)}{m^2} \delta(r_\parallel) \,, \nonumber\\
\hat{e}_4(r)&=& \hat{O}_6(r_\parallel,r_\perp) \hat{O}_2(r_\perp) \dfrac{\tilde{\mathcal{D}}_2(r_\perp)}{m^2} \delta(r_\parallel) \,,
\eea
where
\bea
\hat{O}_3(x, y) &: =&
\frac{x^4-6x^2y^2-2y^4}{6 r^4}   \hat{O}_2(x) - \frac{5 y^2-3r^2}{6 r^4} \left( 2 y^a x^b \dfrac{\partial^2}{\partial y^a \partial x^b} +  y^2 \hat{O}_2(y)\right),\\
\hat{O}_4(x, y) &: =& \frac{ 4 r^4 - 35 y^2 x^2 }{8 r^4}  \hat{O}_2(x)+\frac{5 \left( 3 r^2-7y^2\right) }{4 r^4}  y^a x^b \dfrac{\partial^2}{\partial y^a \partial x^b}+
 \frac{5 y^2\left( 6r^2-7 y^2 \right)-24 r^4 }{8 r^4} \, \hat{O}_2(y) ,\\
\hat{O}_5(x, y) &: =& 
  \frac{7 y^2 x^2}{12 r^4}  \hat{O}_2(x) 
 +\frac{7y^2 - 3 r^2 }{6 r^4}  y^a x^b \dfrac{\partial^2}{\partial y^a \partial x^b} 
- \frac{x^2 \left( r^2 + 7 y^2 \right) }{12 r^4}  \hat{O}_2(y)  ,\\
\hat{O}_6(x, y) &: =&
  \frac{5 y^2 x^2}{4 r^4}  \hat{O}_2(x)
 +\frac{5 y^2 - 3 r^2 }{2 r^4}  y^a x^b \dfrac{\partial^2}{\partial y^a \partial x^b} 
+ \frac{x^2 \left( r^2 -5 y^2 \right) }{4 r^4}  \hat{O}_2(y)\,,
\eea
and 
\bea
\tilde{\mathcal{D}}_i(r_\perp)=\int \dfrac{d^2q_{\perp}}{(2\pi)^2}e^{-i{\bf q}_{\perp}\cdot{\bf r}_{\perp}}\mathcal{D}_i(-{\bf q}_\perp^2)\,.
\eea
The operators $\hat{O}_1$ and $\hat{O}_2$ are defined in Eq.~(\ref{Ooperators}).

\end{document}